\documentclass[aip,amsmath,amssymb,graphicx,reprint]{revtex4-1}

\usepackage{graphicx}
\usepackage{dcolumn}
\usepackage{bm}

\usepackage[utf8]{inputenc}
\usepackage[T1]{fontenc}
\usepackage{mathptmx}

\usepackage{amsmath}
\usepackage{color}
\usepackage{cancel}
\usepackage{ulem}
\usepackage{comment} 
\usepackage{braket}
\usepackage{soul}

\newcommand{\megahz}[1]{{#1}{\ \rm{MHz}}}

\newcommand{\micron}[1]{{#1}{\ \rm{\mu m}}}

\newcommand{\degree}[1]{{#1}^\circ}
\newcommand{\db}[1]{{#1}{\ \rm{dB}}}

\newcommand{\linb}{\rm{LiNbO}_3}
\renewcommand{\vector}[1]{\mathbf{#1}} 

\begin{document}


\title{Optical polarimetric measurement of surface acoustic waves} 



\author{K.~Taga$^{\dagger}$}
\thanks{Author to whom correspondence should be addressed:\\ taga@issp.u-tokyo.ac.jp,\\ hisatomi.ryusuke.2a@kyoto-u.ac.jp,\\ usami@qc.rcast.u-tokyo.ac.jp\\
$^{\dagger}$K.~Taga and R.~Hisatomi contributed equally to this work.}
\affiliation{Research Center for Advanced Science and Technology (RCAST), The University of Tokyo, Meguro-ku, Tokyo 153-8904, Japan}
\author{R.~Hisatomi$^{\dagger}$}
\thanks{Author to whom correspondence should be addressed:\\ taga@issp.u-tokyo.ac.jp,\\ hisatomi.ryusuke.2a@kyoto-u.ac.jp,\\ usami@qc.rcast.u-tokyo.ac.jp\\
$^{\dagger}$K.~Taga and R.~Hisatomi contributed equally to this work.}
\affiliation{Institute for Chemical Research (ICR), Kyoto University, Gokasho, Uji, Kyoto 611-0011, Japan}
\affiliation{PRESTO, Japan Science and Technology Agency, Kawaguchi-shi, Saitama 332-0012, Japan}

\author{Y.~Ohnuma}
\affiliation{Research Center for Advanced Science and Technology (RCAST), The University of Tokyo, Meguro-ku, Tokyo 153-8904, Japan}

\author{R.~Sasaki}
\affiliation{RIKEN Center for Quantum Computing (RQC), RIKEN, Wako, Saitama 351-0198, Japan}

\author{T.~Ono}
\affiliation{Institute for Chemical Research (ICR), Kyoto University, Gokasho, Uji, Kyoto 611-0011, Japan}
\affiliation{Center for Spintronics Research Network, Institute for Chemical
Research, Kyoto University, Gokasho, Uji, Kyoto 611-0011, Japan}
\affiliation{Center for Spintronics Research Network, Graduate School of
Engineering Science, Osaka University, Toyonaka, Osaka 560-0043, Japan}

\author{Y.~Nakamura}
\affiliation{Research Center for Advanced Science and Technology (RCAST), The University of Tokyo, Meguro-ku, Tokyo 153-8904, Japan}
\affiliation{RIKEN Center for Quantum Computing (RQC), RIKEN, Wako, Saitama 351-0198, Japan}

\author{K.~Usami}
\thanks{Author to whom correspondence should be addressed:\\ taga@issp.u-tokyo.ac.jp,\\ hisatomi.ryusuke.2a@kyoto-u.ac.jp,\\ usami@qc.rcast.u-tokyo.ac.jp\\
$^{\dagger}$K.~Taga and R.~Hisatomi contributed equally to this work.}
\affiliation{Research Center for Advanced Science and Technology (RCAST), The University of Tokyo, Meguro-ku, Tokyo 153-8904, Japan}


\date{\today}

\begin{abstract}
Surface acoustic wave (SAW) is utilized in diverse fields ranging from physics, engineering, to biology, for transducing, sensing and processing various signals. Optical measurement of SAW provides valuable information since the amplitude and the phase of the displacement field can be measured locally with the resolution limited by the spot size of the optical beam. So far, optical measurement techniques rely on modulation of optical path, phase, or diffraction associated with SAW. Here, we demonstrate that SAW can be measured with an optical polarimeter. We show that the slope of the periodically tilting surface due to the coherently driven SAW is translated into the angle of polarization rotation, which can be straightforwardly calibrated when polarimeters work in the shot-noise-limited regime. The polarimetric measurement of SAW is thus beneficial for quantitative studies of SAW-based technologies.
\end{abstract}


\maketitle 

Surface acoustic wave (SAW) is a Rayleigh wave propagating on the surface of an elastic media, for which the longitudinal and transverse components are intertwined\cite{LL7,TB2017}. SAWs are ideally suited in transducing, sensing, and processing in a form of coherent phonons, where they can be efficiently and coherently excited and detected electronically through the piezoelectric effect using interdigitated transducers (IDTs). SAWs have thus been used in various electronic devices. SAW filters, for instance, can be ubiquitously found in cellular phones~\cite{R2017,BB2008,C1998}. SAW-based biosensors are now popular in detecting molecules in liquid media~\cite{LR2008,RM2009,ZL2021}. In the emerging field of quantum engineering, SAWs find an opportunity thanks to their long coherence time~\cite{SK2015,MS2018,SZ2018,NY2020,MK2017}. Recently, SAWs are also attracting renewed attention in the field of spintronics, where the ability for transferring their angular momenta to spins could be exploited~\cite{M2013, IMM2014, K2017, K2020,BV2020,WH2012}. 

To expand the possibility of SAW devices, diagnosing spatiotemporal profile of the displacement field of SAW would be beneficial. To this end, various optical measurement techniques are developed. The interferometric method observing the phase modulation due to surface displacement\cite{H2011, K2000,FS2019}, knife-edge method to observe the path modulation due to surface tilt\cite{K2005}, and the diffraction method utilizing the periodic nature of the displacement\cite{H1970}, are explored.    

In this letter, we show that the slope of the periodically tilting surface due to the coherently driven SAW can be sensitively measured with a polarimeter. One of the major motivations to explore the polarimetric measurement of SAW is the fact that the small amount of polarization rotation can be precisely calibrated when polarimeters work in the shot-noise-limited regime. The polarimetric measurement may thus constitute a viable tool to quantitatively evaluate various SAW devices.

\begin{figure*}[t]
\begin{center}
\includegraphics[width=17cm,angle=0]{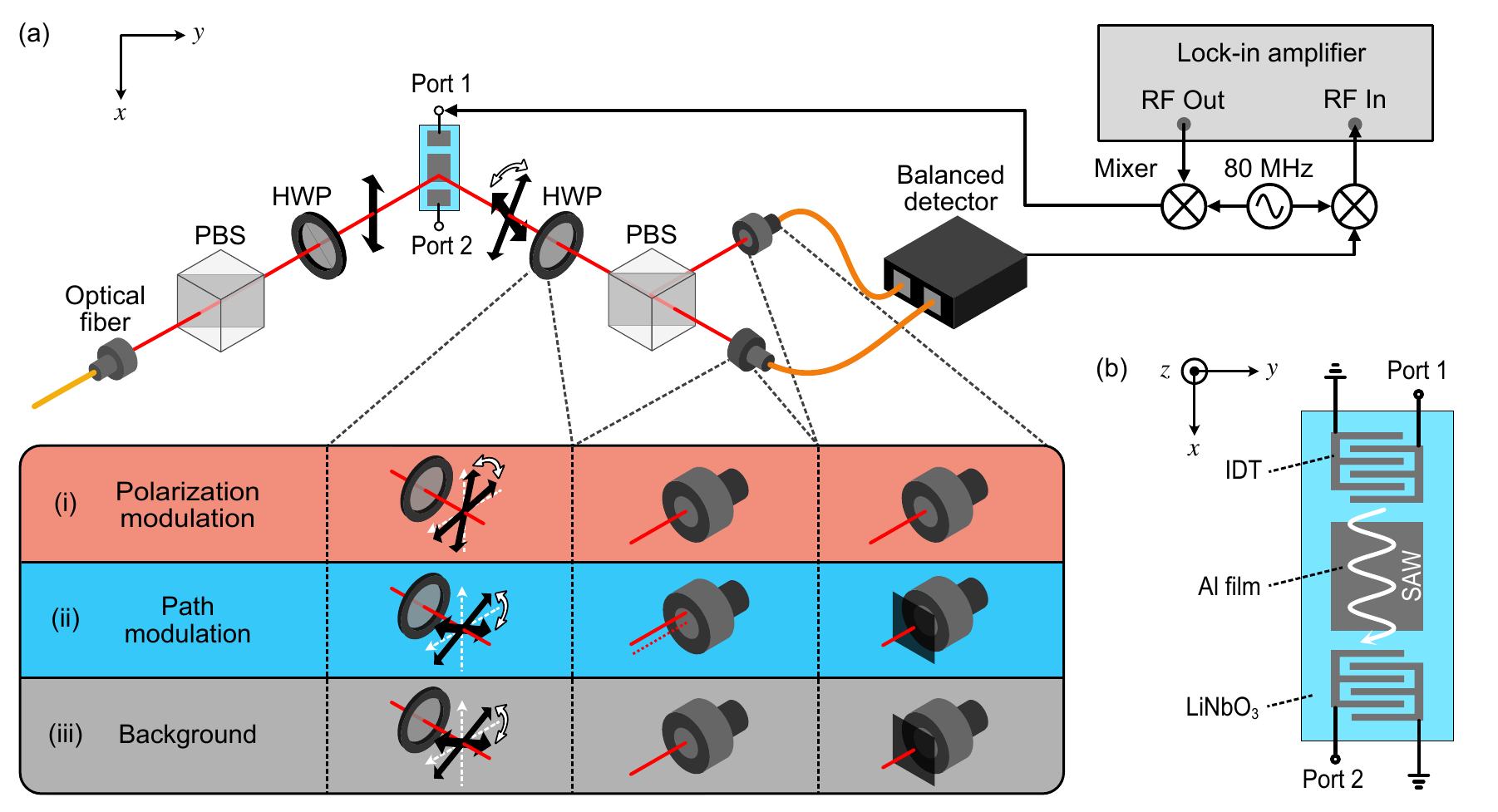}
\caption{
Experimental setup and device for observing optical modulations induced by the SAW. (a)~With a beam splitter (PBS) and a half-wave plate (HWP) the incident light is made linearly polarized with the polarization plane aligned parallel to the direction of SAW propagation ($x$-axis). The beam is focused with an objective lens (not shown) into a diameter of around 10~$\mu$m on the device surface. The angle of incidence is set to $\degree{45}$. The reflected beam is then collimated by another objective lens (not shown) and analyzed in the three different ways indicated as (i), (ii), and (iii) with HWP, PBS, and a balanced detector. The maximum power at the balanced detector is about 70~$\mu$W for each input.
(b)~Schematic illustration of the SAW device with the coordinate system. The device consists of a YZ-cut $\linb$ substrate and aluminum interdigitated transducers (IDTs) which have 50 pairs of 1-mm-long fingers with 10-$\mu$m line and space. The two IDTs are separated by $\micron{812}$, between which a 100-nm-thick Al film with the width of 1280~$\mu$m and the length of 680~$\mu$m is deposited.}
\label{fig:setup}
\end{center}
\end{figure*}

\begin{figure}[t]
\includegraphics[width=8.5cm,angle=0]{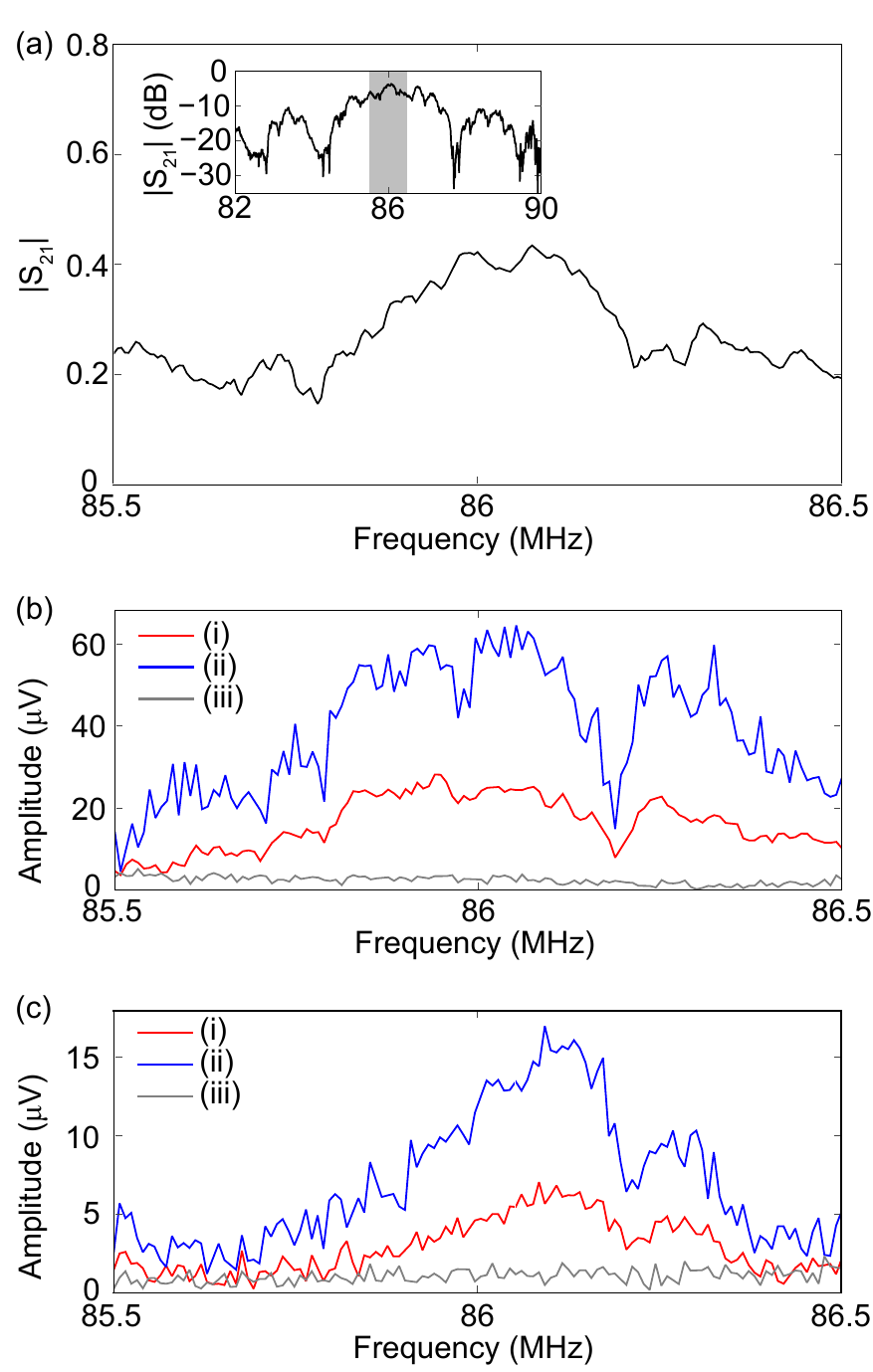}
\caption{Spectra of the SAW propagation, the optical path modulation, and the polarization modulation. (a)~rf transmission spectrum from Port~1 to Port~2. The peak around 86~MHz corresponds to the center frequency of IDTs. The frequency range is the same as that in the following optical modulation spectroscopy. The inset shows the zoom-out of the frequency range in (a) indicated by the gray zone. (b)~Spectra of the polarization modulation amplitude (red), the path modulation signal (blue), and the background signal (gray), each of which is acquired in the corresponding setting denoted as (i)--(iii) in Fig.~\ref{fig:setup}(a). The data is taken in the Al/$\linb$ region. (c)~Spectra taken in the bare $\linb$ region.
}
\label{fig:sweep}
\end{figure}

The experimental setup is schematically shown in Fig.~\ref{fig:setup}(a). We analyze the signal in the three different settings labeled as (i), (ii), and (iii). They are aimed at detecting (i) the polarization modulation, (ii) the optical path modulations, and (iii) the background signal, respectively. 
In the setting (i), the orthogonal polarization components of the reflected beam are split by a PBS, coupled separately to multi-mode fibers, and differentially measured by a balanced detector (Thorlabs:~PDB465C). Here, the polarization plane is rotated with a HWP by the angle of $\degree{45}$ with respect to the original plane in order to maximize the amplitude of the polarization modulation and to eliminate the common-mode intensity fluctuations (see supplementary material). Note that the setting (i) is akin to the one employed to measure the longitudinal magneto-optical Kerr effect~\cite{YS1996}. In the setting~(ii), the polarization plane is rotated by the angle of $\degree{90}$ with respect to the original plane so that one of the split beams has a negligible amplitude while the other maintains the original amplitude. The former beam is then blocked, and the latter beam alone is coupled into the multi-mode fiber with the beam deliberately misaligned along the $x$-axis with respect to the fiber, so that roughly half of the power is fed into the fiber. The setting (ii) is basically the same as the knife-edge method demonstrated before~\cite{K2005}. This setting is insensitive to the polarization modulation but highly sensitive to the deflection of the optical beam path along the $x$-axis (see supplementary material). The setting~(iii) is basically the same as the setting~(ii) but the beam is properly aligned so as to maximize the coupling into the fiber. The background signal (noise floor) measured by this setting is composed of the optical shot noise and the electrical noise of the measurement system.

Figure~\ref{fig:setup}(b) shows the schematic of a SAW device under test. The SAW device is formed by an Al film on a $\linb$ substrate. SAW is excited and detected through interdigitated transducers (IDTs), which convert the radio-frequency~(rf) signal to the SAW signal and vice versa. The SAW with a wavelength of $\sim\micron{40}$ is generated by driving the IDT of Port~1 with an rf signal at the frequency of $\omega_{\mathrm{SAW}}/2 \pi \sim 86$~MHz and propagates along the $x$-direction, that is parallel to the crystalline Z-axis of the $\linb$ substrate. 

We first characterize the SAW propagation in the device. Figure \ref{fig:sweep}(a) shows a transmission spectrum ($|{S}_{21}|$) of the SAW from Port 1 to Port 2 measured with a vector network analyzer. The peak around $\megahz{86}$ corresponds to the center frequency of the band covered by the IDTs. The broad profile appears as a result of the dispersion of the SAW, where the IDTs select preferred wavelengths according to the geometry of the fingers. Small reflections by the IDTs result in the interference and appear as the fine structures in the spectrum. The maximum peak transmission around the center peak ($\sim\db{-4}$) indicates that the rf power loss is sufficiently low even in the presence of the Al film in between. 

Figures~\ref{fig:sweep}(b) and (c) respectively show spectra of the polarization modulations (red curves) and the path modulations (blue curves). Here, the result for the beam spot ($\sim 10$~$\mu$m) placed at a particular positions of the Al/$\linb$ region [bare $\linb$ region] of the device is shown in Fig.~\ref{fig:sweep}(b) [Fig.~\ref{fig:sweep}(c)]. For the measurements, the SAWs are driven from Port 1 with an rf signal with the power of $-10$~dBm produced by mixing a fixed-frequency signal at $\megahz{80}$ from an rf generator (Agilent:~E8247C) and a variable-frequency signal at $\megahz{5.5-6.5}$ from a lock-in amplifier (Zurich Instruments:~HF2LI). The optical signal is acquired by the balanced detector shown in Fig.~\ref{fig:setup}(a), mixed-down with the 80-MHz rf signal, and then demodulated with the lock-in amplifier.
The spectra shown in Figs.~\ref{fig:sweep}(b) and (c) show an enhanced amplitude at around 86~MHz similarly to the rf transmission spectrum shown in Fig.~\ref{fig:sweep}(a). Note that the spectra of the modulated optical signals contain the local information of the SAW and thus sensitive to the interference effects of the SAW. There is no noticeable structure in the spectra of the background signal (gray curves). We confirm that there is no stray rf field directly coupled to the balanced detector, little path modulation signal is found when the beam is properly aligned to maximally couple to the fiber, and little polarization modulation signal is found when the polarization plane of the reflected beam are rotated by the angle of $\degree{90}$ with respect to be the original plane. The spectra of the polarization modulations [red curves in Figs.~\ref{fig:sweep}(b) and~(c)] constitute telltale evidence that the SAW can be measured with the polarimeter. Notably, the signal-to-noise ratio of the polarimetric method is comparable to that of the knife-edge method [blue curves in Figs.~\ref{fig:sweep}(b) and (c)].


\begin{figure*}[t]
\begin{center}
\includegraphics[width=17cm,angle=0]{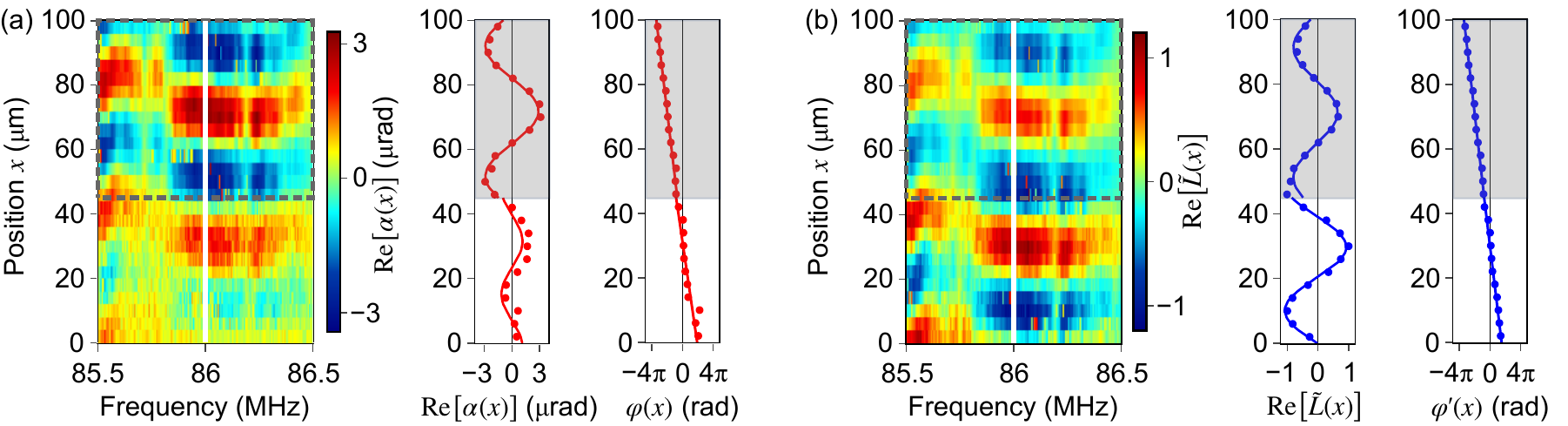}
\caption{Position dependence of the SAW-induced optical modulation signal spectra observed through the polarimetric method (a) and the knife-edge method (b). (a)~Spectra of the polarization modulation $\mathrm{Re}\left[ \alpha(x) \right]$ acquired while the position of the beam spot $x$ is scanned across the boundary between the Al/$\linb$ region (enclosed by the dashed gray rectangle) and the bare $\linb$ region. (b)~Spectra of the normalized path modulation $\mathrm{Re}\left[ \tilde{L}(x) \right]$. Together with the color maps, cross-sections at the frequency of $\megahz{86}$ (indicated by white lines in the color maps) and their associated phases are shown with fitting curves.}
\label{fig:mapping}
\end{center}
\end{figure*}

To gain further insight the spectra of the amplitude and the phase are acquired with respect to the drive frequency $\omega_{\mathrm{SAW}}/2\pi$ while the position of the beam spot is scanned by $\micron{100}$ across the border between the Al/$\linb$ region and the bare $\linb$ region. Figures~\ref{fig:mapping}(a) and (b) show the spectra of the in-phase (real) component of the polarization modulation amplitude $\mathrm{Re}\left[\alpha(x)\right]$ and that for the path modulation amplitudes $\mathrm{Re}\left[L(x)\right]$ together with the respective cross-sections at the frequency of $\megahz{86}$ and the phases $\varphi(x)$ and $\varphi'(x)$. Here, the polarization modulation amplitude $\alpha(x)$ is calibrated with the shot noise (see supplementary material) and denoted in terms of rotation angle in units of radian. On the other hand, the amplitude $L(x)$ is not easily calibrated in units of meter and is thus denoted as the dimensionless value $\tilde{L}(x)$, which is normalized with respect to the optical power reflected off the surface position probed~(see supplementary material). These procedures enable us to make a fair comparison between the signals in the highly-reflective Al/$\linb$ region and the modestly-reflective bare $\linb$ region. 

As shown in Figs.~\ref{fig:mapping}(a) and (b), the phases $\varphi(x)$ and $\varphi'(x)$ linearly develop as a function of the probe position $x$, i.e., $\varphi(x) = q_{\mathrm{Al}} x$ ($q_{\mathrm{LN}} x$) and $\varphi'(x) = q'_{\mathrm{Al}} x$ ($q'_{\mathrm{LN}} x$), in the Al/$\linb$ region (the bare $\linb$ region), respectively. Here, we obtain the wavelengths $\lambda_{\mathrm{Al}} \equiv \frac{2\pi}{q_{\mathrm{Al}}} = 40.3 \pm 0.6$ ($32.7 \pm 4.0$)~$\mu$m in the Al/$\linb$ region (the bare $\linb$ region) from the polarization modulation signal and $\lambda'_{\mathrm{Al}} \equiv \frac{2\pi}{q'_{\mathrm{Al}}} = 41.7 \pm 1.1$ ($39.5 \pm 0.5$)~$\mu$m in the Al/$\linb$ region (the bare $\linb$ region) from the path modulation signal. With separate experiments~(see supplementary material), we verify that the phases $\varphi(x)$ and $\varphi'(x)$ develop oppositely when the propagating direction of the SAW is switched. Assuming that the polarization-modulation amplitudes are uniform within the Al/$\linb$ region, we obtain $\alpha_{\mathrm{Al}}=2.91 \pm 0.09\ (2.82 \pm 0.12)~\mu$rad by fitting the sinusoidal signal $\mathrm{Re}\left[\alpha(x)\right]$ in the middle panel of Fig.~\ref{fig:mapping}(a) (Fig.~S3(a)) with $\alpha_{\mathrm{Al}} \cos(q_{\mathrm{Al}} x)$, when the SAW propagates in the $+x$ ($-x$)-direction. Similar fitting with $\alpha_{\mathrm{LN}} \cos(q_{\mathrm{LN}} x)$ yields $\alpha_{\mathrm{LN}} = 1.15 \pm 0.34\ (1.22 \pm 0.17)~\mu$rad for the spot position $x$ at the bare LiNbO$_3$ region.

\begin{figure*}[t]
\begin{center}
\includegraphics[width=17cm,angle=0]{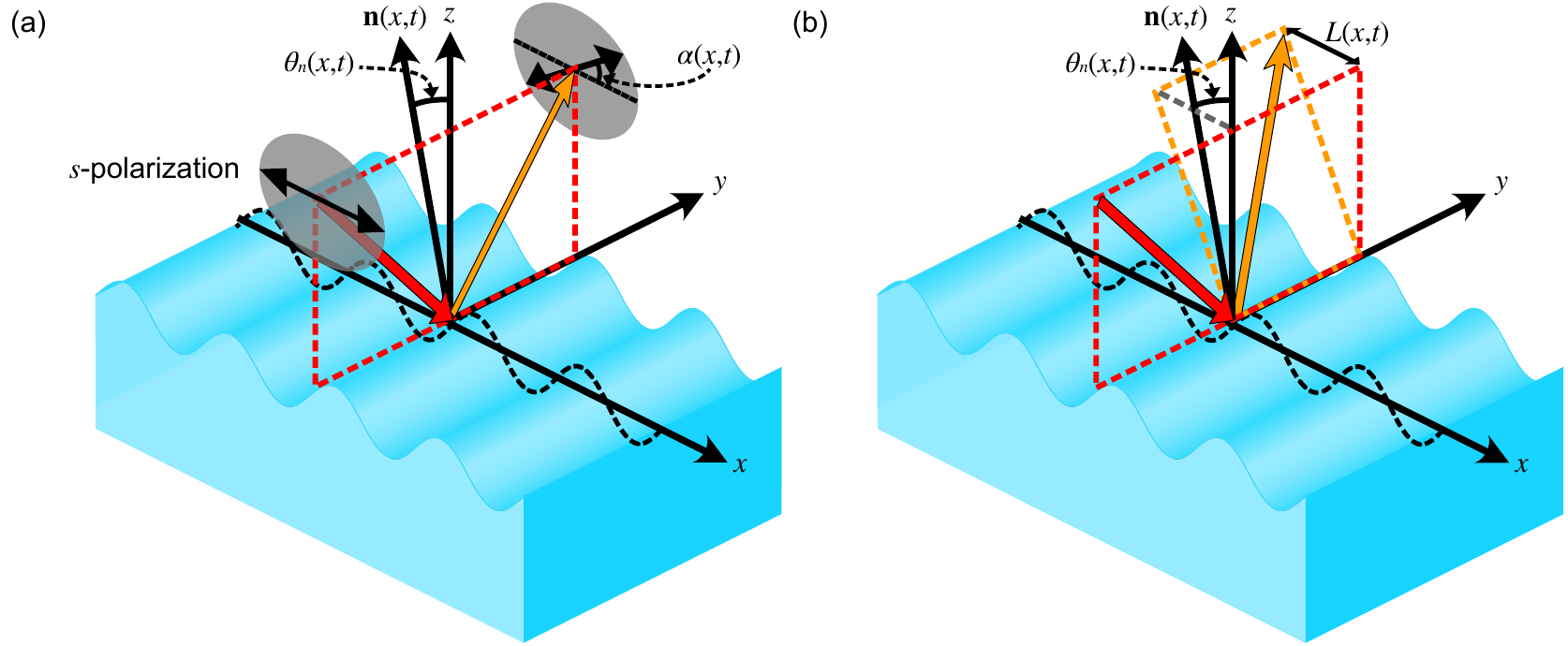}
\caption{Schematic illustrations of (a) the polarization modulation and (b) the path modulation. Variables $\vector{n}(x,t)$, $\theta_{n}(x,t)$, $\alpha(x,t)$, and $L(x,t)$ represent the normal vector of the surface at the light spot, the slope angle of the surface, the angle of the polarization rotation, and the departure of the reflected light from the nominal position, respectively.}
\label{fig:trajectory}
\end{center}
\end{figure*}

The observed polarization and path modulations of the reflected light can be considered as the results of the moving boundary effect due to the SAW as illustrated in Fig.~\ref{fig:trajectory}. Here, the boundary is the surface of the device having the normal vector $\vector{n}(x,t)$, which is periodically tilted with the slope angle denoted by $\theta_{n}(x,t) = \frac{\partial u_{z}(x,t)}{\partial x}$. Here, $u_{z}(x,t)$ is the $z$-component of displacement field at position $x$ and time $t$ associated with the SAW. In the experimental configuration depicted in Fig.~\ref{fig:setup}(a), the tilting surface with the slope $\theta_{n}(x,t)$ results in the polarization rotation in such a way as
\begin{equation}
\alpha (x,t) = \mathrm{Re}\left[\alpha(x) e^{-i \omega_{\mathrm{SAW}}t}\right] = \sqrt{2}\left(\frac{r_{p}}{r_{s}}-1 \right)\theta_{n}(x,t), \label{eq:pol}
\end{equation} 
where $r_{p}$ and $r_{s}$ are the Fresnel reflection coefficients for s-polarization and p-polarization, respectively (see supplementary material). The amplitude of the path modulation $L(x,t)$ can also be expressed in terms of the slope angle as
\begin{equation}
L(x,t) = \mathrm{Re}\left[L(x) e^{-i \omega_{\mathrm{SAW}}t}\right] = 2l_{0}\theta_{n}(x,t), \label{eq:path}
\end{equation}
where $l_{0}$ having the dimension of length is the proportional factor. It is then evident that the phases of $\mathrm{Re}\left[\alpha(x)\right]$ and $\mathrm{Re}\left[L(x)\right]$ may develop along $x$ in the same way since both are dictated by $\theta_{n}(x,t)$. This is consistent with the observation shown in Fig.~\ref{fig:mapping}. In separate experiments (see supplementary material), we also confirm that the modulation amplitudes both depend linearly on the amplitude of the slope $\theta_{n}(x,t)$, and thus $z$-component of the displacement field $u_{z}(x,t)$ (which is proportional to the square root of the rf power used for driving the SAW). These facts support the model in which the polarization modulation is directly linked to the slope angle $\theta_{n}(x,t)$.

One of the most appealing features of the polarimetric measurement is the fact that the amount of polarization rotation can be precisely calibrated with shot noise when working in the shot-noise-limited regime (see supplementary material). Given the polarization rotation angle, the slope angle $\theta_{n}(x)$ can be estimated from Eq.~(\ref{eq:pol}). We have $\theta_{n,\mathrm{Al}}= 1.04 \pm 0.03\ (1.00 \pm 0.04)~\mu$rad for $x$ in the Al/LiNbO$_3$ region and $\theta_{n,\mathrm{LN}} = 0.54 \pm 0.16\ (0.58 \pm 0.08)~\mu$rad for $x$ in the bare LiNbO$_3$ region, when the SAW propagates in the $+x$ ($-x$)-direction, which are in reasonable agreement with the value $\left|\theta_{n}(x)\right|\sim 1.8~\mu$rad separately estimated from the physical dimension of the device and the driving power (see supplementary material). There is a discrepancy between $\theta_{n,\mathrm{Al}}$ and $\theta_{n,\mathrm{LN}}$, though. The discrepancy may be caused by the difference between the amplitudes of the displacement $\left|u_{z}(x) \right|$ in the Al/LiNbO$_{3}$ region and that in the LiNbO$_{3}$ region due to, e.g., extra reflection at the boundary between the Al/LiNbO$_{3}$ region and the bare LiNbO$_3$ region, and the resultant interference effect. Other possibilities may be due to the optoelastic effect or the optical uniaxial anisotropy of the LiNbO$_{3}$ single crystal.

Finally, we note that, from the calibrated polarization-modulation amplitude, we can negate another mechanism that might cause polarization rotation, namely, the magneto-optical Kerr effect. The Kerr effect here could be caused by the effective magnetic field created by the SAW-induced vortex field~\cite{M2013,IMM2014,K2017,K2020}. As discussed in supplementary material, the expected Kerr rotation is too small to account for the observed polarization rotation.

In summary, we demonstrated the polarimetric measurement of SAW. The signal-to-noise ratio of the polarimetric method is on per with that of the knife-edge method. The precisely calibrated polarimetric measurement can be a viable tool to quantitatively evaluate the spatiotemporal profile of the displacement field of SAW.


\vspace{1cm}
\textbf{Acknowledgements} We are indebted to Rekishu Yamazaki, Atsushi Noguchi,  Yosuke Nakata, Maria Fuwa, and Shotaro Kano, for useful discussion. We acknowledge financial support from JST ERATO (Grant Number JPMJER1601), JSPS KAKENHI (Grant Number JP20K15162), JST PRESTO (Grant Number MJPR200A), and the Collaborative Research Program of the Institute for Chemical Research, Kyoto University.

\vspace{1cm}
\textbf{Author contributions} K.~T., R.~H., T.~O., and K.~U. conceived the experiment. K.~T. and R.~H. built the experimental setup, fabricated the SAW device, and performed the measurements. K.~T., R.~H., Y.~O., R.~S., and K.~U. analysed the data and developed the phenomenological model. With the inputs from Y.~O, T.~O., and Y.~N., K.~T., R.~H., and K.~U. wrote the manuscript. T.~O., Y.~N., and K.~U. provided general research supervision.

\vspace{1cm}
\textbf{Data Availability} The data that support the findings of this study are available
from the corresponding author upon reasonable request.

\appendix

\begin{widetext}



\section{T\lowercase{heoretical models of the polarization and path modulations}}
The observed polarization and path modulations of the reflected light can be considered as the results of the moving boundary effect due to the SAW. Let us suppose that without driving the SAW the wavevectors of the incident light and the reflected light are both in the $yz$-plane as shown in Fig.~4. When we drive the SAW, the normal vector $\vector{n}(x,t)$ of the surface of the device (boundary between the device and the air) at the optical spot becomes position- as well as time-dependent and is periodically tilting in space and time. The periodically tilting normal vector $\vector{n}(x,t)$ in turn leads to the polarization and path modulations. Now, we shall describe how these two modulations occur in more detail.

\subsection{Polarization modulation}\label{sec:polarization}
To simplify the discussion, let us suppose that the wavevectors of the incident light and the reflected light are both in the $yz$-plane as shown in Fig.~4(a) in the main text even though the reflected light departs from the $yz$-plane from time to time as a result of path modulation as will be described in Sec.~\ref{sec:path}. The assumption is justified in view of our particular experimental configuration in measuring the polarization modulation. The justification of the assumption will be reexamined in Sec.~\ref{sec:measurements}.

Without driving the SAW, the relationship between the electric fields of the incident and reflected light is simply determined by the \textit{Fresnel reflection formula}:
\begin{equation}
\left[
\begin{array}{c}
a_{p}^{r}(t)  \\
a_{s}^{r}(t) 
\end{array}
\right] = \left[
\begin{array}{cc}
r_p & 0 \\
0 & r_s \\
\end{array}
\right] 
\left[
\begin{array}{c}
a_{p}^{i}(t)  \\
a_{s}^{i}(t) 
\end{array}
\right],\label{eq:Fresnel0}
\end{equation}
where $a_p^{i}(t)$ and $a_s^{i}(t)$ are the dimensionless electric fields for $p$- and $s$-polarized incident light such that, with $\omega_{0}$ being the angular frequency of the incident light, the power of incident light for the $p$- and $s$-polarized components  are given by $\hbar \omega_{0} \left(a_p^{i\ *}(t) a_p^{i}(t) \right)$ and $\hbar \omega_{0} \left( a_s^{i\ *}(t) a_s^{i}(t) \right)$, respectively. $a_p^{r}(t)$ and $a_s^{r}(t)$ are the dimensionless electric fields for the reflected light that are similarly defined as for the incident light. The Fresnel reflection coefficients $r_p$ for the $p$-polarized light and $r_s$ for the $s$-polarized light are given by
\begin{eqnarray}
r_{p} &=& \frac{n_{2} \cos \theta_{1} - n_{1} \cos \theta_{2} }{n_{1} \cos \theta_{2} + n_{2} \cos \theta_{1}} \label{eq:rp}\\
r_{s} &=& \frac{n_{1} \cos \theta_{1} - n_{2} \cos \theta_{2} }{n_{1} \cos \theta_{1} + n_{2} \cos \theta_{2}}, \label{eq:rs}
\end{eqnarray}
respectively. Here, $n_{1}\left(=1\right)$ and $n_{2}$ are the complex refractive indices of the air and the device and $\theta_{1}\left(=\frac{\pi}{4}\right)$ and $\theta_{2}$ are the incident and refracted angles, respectively. Note that $\theta_{2}$ can be determined by the Snell's law:
\begin{equation}
n_{1} \sin \theta_{1} = n_{2} \sin \theta_{2}.
\end{equation}
Since $n_{2}$ is in general a complex number, $\theta_{2}$ as well as $r_{p}$ and $r_{s}$ are in general complex numbers.

By driving the SAW, the normal vector $\vector{n}(x,t)$, periodically tilts from the $z$-axis in the $xz$-plane by the amount of $\theta_{n}(x,t)$, which is given by
\begin{equation}
\theta_n(x,t) = \frac{\partial u_{z}(x, t)}{\partial x}, \label{eq:theta_n}
\end{equation}
where 
\begin{equation} 
u_{z}(x,t)\equiv \mathrm{Re}\left[ u_{z}(x) e^{-i\omega_{\mathrm{SAW}}t}\right]
\end{equation}
is the $z$ component of the displacement field at position $x$ and time $t$ due to the coherently driven SAW at the angular frequency $\omega_{\mathrm{SAW}}$. Here, the $x$-dependent complex amplitude $u_{z}(x)$ can also be written as
\begin{equation}
u_{z}(x)=\left|u_{z}(x)\right|e^{i\varphi(x)} \label{eq:u_z}
\end{equation}
with $\varphi(x)$ being the $x$-dependent phase. Note that $\left|u_{z}(x)\right|$ can be considered to be more or less constant and the spatial variation of $u_{z}(x)$ is encoded in the phase factor $\varphi(x) = q x$, where $q=\frac{2\pi}{\lambda}$ is the wave vector of the SAW with $\lambda$ being the wavelength. With $u_{z}(x)$, the slope angle $\theta_{n}(x,t)$ in Eq.~(\ref{eq:theta_n}) can be written as
\begin{equation}
\theta_{n}(x,t) =\mathrm{Re}\left[ \underbrace{\frac{\partial u_{z}(x)}{\partial x}}_{\theta_{n}(x)} e^{-i \omega_{\mathrm{SAW}}t} \right], \label{eq:theta_n2}
\end{equation}
where 
\begin{equation}
\theta_{n}(x) = \left|\theta_{n}(x)\right| e^{i\varphi(x)} \label{eq:theta_n3}
\end{equation} 
is the $x$-dependent complex amplitude of $\theta_{n}(x,t)$. Note that $\left|\theta_{n}(x)\right|$ is of the order of $q\left|u_{z}(x)\right| \sim \frac{\left|u_{z}(x)\right|}{\lambda}$. Since $\lambda \sim 40$~$\mu$m and $\left|u_{z}(x)\right|\sim 10$~pm in our experiments, we shall henceforth assume that $\left|\theta_{n}(x)\right| \ll 1$.

\begin{figure*}[h]
\begin{center}
\includegraphics[width=12cm,angle=0]{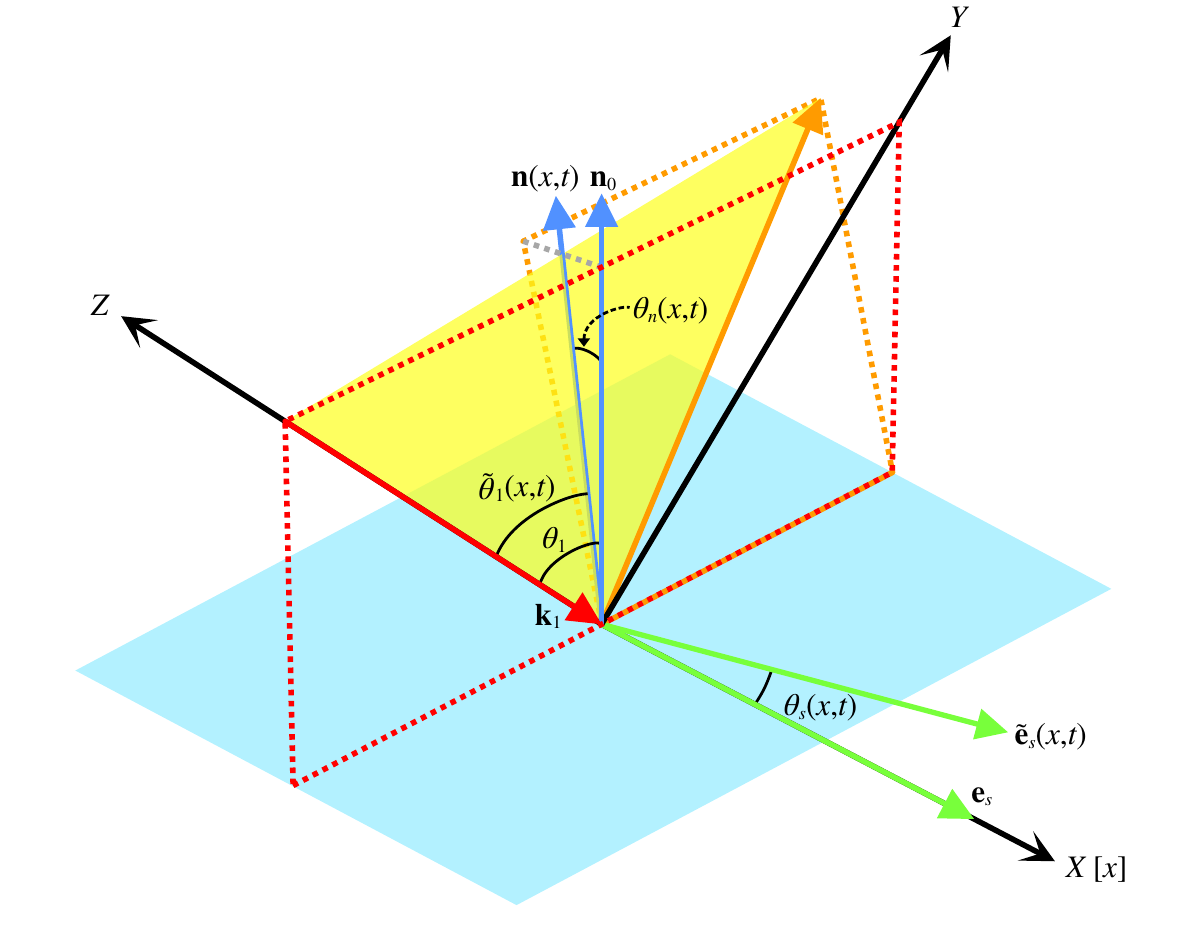}
\caption{Schematic illustrations of the change of the plane of incidence. The nominal incidence angle $\theta_{1}$ is the subtended angle between the nominal normal vector, $\vector{n}_{0}=\left[\begin{array}{c}
0 \\
\sin{\theta_{1}} \\
\cos{\theta_{1}}
\end{array}\right]$, and an unit vector parallel to the wavevector of the incident light $\vector{k}_{1}=\left[\begin{array}{c}
0 \\
0 \\
-1
\end{array}\right]$. The nominal incident angle $\theta_{1}$ changes into $\tilde{\theta}_{1}(x,t)$ when we drive the SAW. $\tilde{\theta}_{1}(x,t)$ is defined as the subtended angle between the normal vector $\vector{n}(x,t)=\left[\begin{array}{c}
\sin{\theta_{n}(x,t)} \\
\cos{\theta_{n}(x,t)}\sin{\theta_{1}} \\
\cos{\theta_{n}(x,t)}\cos{\theta_{1}}
\end{array}\right]$ and the wavevector of the incident light $\vector{k}_{1}$. Furthermore, the nominal basis vector for the $s$-polarization $\vector{e}_{s}=\left(\vector{k}_{1} \times \vector{n}_{0}\right)/ \left|\vector{k}_{1} \times \vector{n}_{0}\right| =\left[\begin{array}{c}
1 \\
0 \\
0
\end{array}\right]$ changes into $\tilde{\vector{e}}_{s}(x,t)=\left(\vector{k}_{1} \times \vector{n}(x,t)\right)/ \left|\vector{k}_{1} \times \vector{n}(x,t)\right|=\frac{1}{\sqrt{\left(\cos{\theta_{n}(x,t)} \sin{\theta_{1}}\right)^{2} + \sin{\theta_{n}(x,t)}}}\left[\begin{array}{c}
\cos{\theta_{n}(x,t)} \sin{\theta_{1}}\\
-\sin{\theta_{n}(x,t)}  \\
0
\end{array}\right]$ when we drive the SAW. The subtended angle $\theta_{s}(x,t)$ between $\vector{e}_{s}$ and $\tilde{\vector{e}}_{s}(x,t)$ defines the rotation angle of incident plane about $\vector{k}_{1}$. The angle $\theta_{s}(x,t)$ can be given by the following relation: $\cos{\theta_{s}(x,t)}=\vector{e}_{s} \cdot \tilde{\vector{e}}_{s}(x,t) = \frac{\sin{\theta_{1}}}{\sqrt{ \sin^{2}\theta_{1} + \tan^{2} \theta_{n}(x,t)}}$. Note that the coordinate system denoted by $X$, $Y$, and $Z$ is different from the one used in Fig.~4, while $X$ is equivalent to $x$.}
\label{fig:S1}
\end{center}
\end{figure*}

When the SAW is driven, the formula Eq.~(\ref{eq:Fresnel0}) is modified in two ways. One is that the incident angle $\theta_{1}$ changes into $\tilde{\theta}_{1}(x,t)$ because the plane of incidence changes as the normal vector $\vector{n}(x,t)$ changes due to the SAW. The incident angle $\tilde{\theta}_{1}(x,t)$ is the angle subtended by the normal vector $\vector{n}(x,t)=\left[ 
\begin{array}{c}
\sin \theta_{n}(x,t) \\
\cos \theta_{n}(x,t) \sin \theta_{1} \\
\cos \theta_{n}(x,t) \cos \theta_{1}
\end{array}
\right]$ and the wavevector of the incident light $\vector{k}_{1} = \left[
\begin{array}{c}
0 \\
0 \\
1
\end{array}
\right]$ as shown in Fig.~\ref{fig:S1}. Here, the different coordinates denoted by $Y$ and $Z$ are used as opposed to the coordinates $y$ and $z$ in Fig.~1(a). We have thus the following relation:
\begin{equation}
\cos \tilde{\theta}_{1}(x,t) = \vector{n}(x,t) \cdot \vector{k}_{1} = \cos \theta_{1} \cos \theta_{n}(x,t).
\end{equation}
With the instantaneous position-dependent value of $\tilde{\theta}_{1}(x,t)$, we have 
\begin{eqnarray}
\tilde{r}_{p}(x,t) &=& \frac{n_{2} \cos \tilde{\theta}_{1}(x,t) - n_{1} \cos \tilde{\theta}_{2}(x,t)}{n_{1} \cos \tilde{\theta}_{2}(x,t) + n_{2} \cos \tilde{\theta}_{1}(x,t) } \label{eq:rp_2}\\
\tilde{r}_{s}(x,t) &=& \frac{n_{1} \cos \tilde{\theta}_{1}(x,t) - n_{2} \cos \tilde{\theta}_{2}(x,t) }{n_{1} \cos \tilde{\theta}_{1}(x,t) + n_{2} \cos \tilde{\theta}_{2}(x,t) }, \label{eq:rs_2}
\end{eqnarray}
where $\tilde{\theta}_{2}(x,t)$ is determined by the Snell's law
\begin{equation}
n_{1} \sin \tilde{\theta}_{1}(x,t) = n_{2} \sin \tilde{\theta}_{2}(x,t).
\end{equation}
Note that for $\theta_{n}(x,t)\ll 1$ we have $\tilde{\theta}_{1} \sim \theta_{1}$ and Eqs.~(\ref{eq:rp_2}) and (\ref{eq:rs_2}) reduce to Eqs.~(\ref{eq:rp}) and (\ref{eq:rs}), respectively.

The other is that the eigenbasis for the vectors $\left[\begin{array}{c}
a_{p}^{i}(t) \\
a_{s}^{i}(t)
\end{array} \right]$ and $\left[\begin{array}{c}
a_{p}^{r}(t) \\
a_{s}^{r}(t)
\end{array} \right]$ in Eq.~(\ref{eq:Fresnel0}) changes as the plane of incidence changes due to the SAW. The change of the eigenbasis can be represented by a rotation matrix 
\begin{equation}
R(\theta_{s}(x,t)) = \left[
\begin{array}{cc}
\cos \theta_{s}(x,t)  & -\sin \theta_{s}(x,t)  \\
\sin \theta_{s}(x,t)  & \cos \theta_{s}(x,t)\\
\end{array}
\right] \label{eq:R}
\end{equation} 
that mixes $p$-polarized and $s$-polarized components of light. Here, the axis of the rotation represented by Eq.~(\ref{eq:R}) is along the wavevector of the incident light $\vector{k}_{1}$. From the geometrical consideration (see Fig.~\ref{fig:S1}), the rotation angle $\theta_{s}(x,t)$ about $\vector{k}_{1}$ can be given by
\begin{equation}
\cos \theta_s(x,t) = \frac{\sin \theta_{1}}{\sqrt{\sin^{2} \theta_{1} + \tan^{2} \theta_{n}(x,t)}},
\end{equation} For the case where the incident angle $\theta_{1} = \frac{\pi}{4}$ and $\theta_{n}(x,t) \ll 1$, we have $\theta_{s}(x,t) = \sqrt{2} \theta_{n}(x,t)$.

By putting everything together, Eq.~(\ref{eq:Fresnel0}) becomes, under the assumption of $\theta_{1} = \frac{\pi}{4}$ and $\theta_{n}(x,t) \ll 1$, 
\begin{equation}
\left[
\begin{array}{c}
a_{p}^{r}(t)  \\
a_{s}^{r}(t) 
\end{array}
\right] = R(-\sqrt{2}\theta_n(x,t))\left[
\begin{array}{cc}
r_p & 0 \\
0 & r_s \\
\end{array}
\right] 
R(\sqrt{2}\theta_n(x,t))
\left[
\begin{array}{c}
a_{p}^{i}(t)  \\
a_{s}^{i}(t) 
\end{array}
\right] = \left[
\begin{array}{cc}
A(x,t) & C(x,t) \\
C(x,t) & B(x,t)
\end{array}
\right]\left[
\begin{array}{c}
a_{p}^{i}(t)  \\
a_{s}^{i}(t) 
\end{array}
\right],\label{eq:Fresnel}
\end{equation}
where
\begin{eqnarray}
A(x,t) &=& r_{p} \cos^{2} \left(\sqrt{2}\theta_{n}(x,t)\right) + r_{s} \sin^{2} \left(\sqrt{2}\theta_{n}(x,t)\right) \label{eq:A}\\
B(x,t) &=& r_{p} \sin^{2} \left(\sqrt{2}\theta_{n}(x,t)\right) + r_{s} \cos^{2} \left(\sqrt{2}\theta_{n}(x,t)\right) \label{eq:B}\\
C(x,t) &=& -\left( r_{p} - r_{s} \right) \cos \left(\sqrt{2}\theta_{n}(x,t)\right) \sin \left(\sqrt{2}\theta_{n}(x,t)\right). \label{eq:C}
\end{eqnarray}
When $\theta_n(x,t)$ is zero as in Eq.~(\ref{eq:Fresnel0}), the matrix relating the incident electric field to the reflected electric field has only diagonal components, but when $\theta_n(x,t)$ is non-zero, off-diagonal components appear as in Eq.~(\ref{eq:Fresnel}). This is the origin of the polarization rotation.

From Eq.~(\ref{eq:Fresnel}), if the input light is $s$-polarized as in the experiments, the electric field of the reflected light is written as
\begin{equation}
\left[
\begin{array}{c}
a_{p}^{r}(t)  \\
a_{s}^{r}(t)
\end{array}
\right] = \left[
\begin{array}{cc}
C(x,t) a_{s}^{i}(t) \\
B(x,t) a_{s}^{i}(t) \\
\end{array}
\right] 
\sim
\left[
\begin{array}{cc}
-\sqrt{2}(r_p - r_s)\theta_n(x,t) \\
r_s \\
\end{array}
\right] 
a_{s}^{i}(t).\label{eq:Fresnel2}
\end{equation}
Note that since $\theta_n(x,t)$ is usually very small, we ignore the higher-order terms above the second order in $\theta_n(x,t)$.
The polarization rotation angle $\alpha(x,t)$ can then be written as
\begin{eqnarray}
\alpha(x,t) \sim \tan(\alpha(x,t)) &=& -\frac{a_p^r(t)}{a_s^r(t)} \nonumber \\
&=& \sqrt{2}\left(\frac{r_p}{r_s}-1\right)\theta_n(x,t) = \mathrm{Re}\left[\underbrace{\sqrt{2}\left(\frac{r_p}{r_s}-1\right)\theta_n(x)}_{\alpha(x)} e^{-i \omega_{\mathrm{SAW}}t}\right]. \label{eq:rotation_angle}
\end{eqnarray}
Note that, in the lock-in detection with the reference being the SAW drive signal having the angular frequency of $\omega_{\mathrm{SAW}}$, we have $\mathrm{Re}\left[\alpha(x)\right]$ for the in-phase detection and $\mathrm{Im}\left[\alpha(x)\right]$ for the quadrature detection. Since the ratio $\frac{r_p}{r_s}$ appears in Eq.~(\ref{eq:rotation_angle}) is a material-specific value, the rotation angle $\alpha(x,t)$ for given $\theta_n(x,t)$ is different from material to material.

\subsection{Path modulation} \label{sec:path}
To be complete, let us discuss how the periodically tilting normal vector $\vector{n}(x,t)$ leads to the path modulation. Suppose that the spot diameter of the incident light is sufficiently small so that Snell's law holds locally at the boundary between the device and the air. The normal vector $\vector{n}(x,t)$ of the device surface oscillates within the $xz$-plane both in position $x$ and time $t$ due to the coherently driven SAW. Consequently, the wavevector of the reflected light also periodically departs from the $yz$-plane as shown in Fig.~4(b) in the main text. The path of the reflected light projected onto the $xz$-plane then varies in accordance with the normal vector $\vector{n}(x,t)$ at position $x$ and time $t$.
This leads to the path modulation of the reflected light.
Given the angle $\theta_n(x,t)$ between the normal vector $\vector{n}(x,t)$ and the $z$-axis as defined by Eq.~(\ref{eq:theta_n2}), the departure of the reflected light from the nominal position can be written as
\begin{eqnarray}
L(x,t) &\equiv& l_{0}\tan\left(2\theta_{n}(x,t)\right) \nonumber \\
&\sim& 2l_{0} \theta_{n}(x,t)= \mathrm{Re} \left[ \underbrace{2 l_{0} \theta_{n}(x)}_{L(x)} e^{-i \omega_{\mathrm{SAW}}t} \right], \label{eq:dl}
\end{eqnarray}
where $l_{0}$ is introduced as a proportional constant that depends on the length from the reflection surface to the observer. Note that, in the lock-in detection with the reference being the SAW drive signal having the angular frequency of $\omega_{\mathrm{SAW}}$, we have signals proportional to $\mathrm{Re}\left[L(x)\right]$ for the in-phase detection and $\mathrm{Im}\left[L(x)\right]$ for the quadrature detection.

\section{D\lowercase{etails of the measurement settings}} \label{sec:measurements}
The reflected light interacting with the SAW on the sample surface is modulated in polarization and path simultaneously. In this section, we describe how the the polarization and the path modulations can be separately measured in the settings (i) and (ii) in Fig.~1(a), respectively.

\subsection{Polarization  modulation}\label{sec:pol_mod}
When observing the polarization modulation in the setting (i), what we actually measure is the intensity difference obtained with the balanced detector as shown in Fig.~\ref{fig:setup_zoom}. The output voltage of the balanced detector is proportional to the difference between the $X$- and $Z$-polarized photon fluxes. Here, $X$- and $Z$-polarization states after the half wave plate (HWP) in Fig.~\ref{fig:setup_zoom} correspond to $\frac{\pi}{4}$- and $-\frac{\pi}{4}$-polarization states before the HWP. Thus, referring to the polarization states before the HWP in Fig.~\ref{fig:setup_zoom}, the observable that the balanced detector measures is the Stokes parameter $S_{X}^{r}(t)$ for the reflected light, which is defined as 
\begin{eqnarray}
S_{X}^{r}(t) &=& \frac{1}{2}\left( a_{\frac{\pi}{4}}^{r\ *}(t)  a_{\frac{\pi}{4}}^{r}(t)- a_{-\frac{\pi}{4}}^{r \ *}(t) a_{-\frac{\pi}{4}}^{r}(t) \right) \nonumber \\
&=& \frac{1}{2}\left( a_{p}^{r\ *}(t) a_{s}^{r}(t)+ a_{s}^{r\ *}(t)a_{p}^{r}(t)\right), \label{eq:obs}
\end{eqnarray}
where in the second line we use the following relations, $a_{\frac{\pi}{4}}^{r}(t) =\frac{1}{\sqrt{2}}\left( a_{p}^{r}(t)+a_{s}^{r}(t) \right)$ and $a_{-\frac{\pi}{4}}^{r}(t)=\frac{1}{\sqrt{2}}\left( a_{p}^{r}(t)-a_{s}^{r}(t) \right)$. 

We now show that the observable $S_{X}^{r}(t)$ contains the information of the slope angle $\theta_{n}(x,t)$ through the angle of the polarization rotation $\alpha(x,t)$ defined by Eq.~(\ref{eq:rotation_angle}). To this end, let us first rewrite Eq.~(\ref{eq:Fresnel}) in terms of the corresponding Stokes parameters:
\begin{eqnarray}
\left[
\begin{array}{c}
S_{X}^{r}(t) \\
S_{Y}^{r}(t) \\
S_{Z}^{r}(t) 
\end{array}
\right] &=& \left[
\begin{array}{ccc}
A(x,t)B(x,t)+C(x,t)^{2} & 0 & \left(A(x,t)-B(x,t) \right) C(x,t) \\
0 & A(x,t)B(x,t)-C(x,t)^{2} & 0 \\
\left(A(x,t)-B(x,t) \right) C(x,t) & 0 & \frac{A(x,t)^{2}+B(x,t)^{2}}{2} - C(x,t)^{2} \\
\end{array}\right]\left[
\begin{array}{c}
S_{X}^{i}(t) \\
S_{Y}^{i}(t) \\
S_{Z}^{i}(t)
\end{array}
\right] \nonumber \\
&\ & \ \ \ + \left[
\begin{array}{c}
\left(A(x,t)+B(x,t)\right) C(x,t) \\
0 \\
\frac{A(x,t)^{2}-B(x,t)^{2}}{2}
\end{array}
\right] S_{0}^{i}(t), \label{eq:affine_map}
\end{eqnarray}
where the Stokes parameter $S_{X}^{i}(t)$ for the incident light is defined as
\begin{equation}
S_{X}^{i}(t) = \frac{1}{2}(a_{p}^{i\ *}(t) a_{s}^{i}(t)+ a_{s}^{i\ *}(t)a_{p}^{i}(t)),
\end{equation}
and the other Stokes parameters are defined as
\begin{eqnarray}
S_{Y}^{k}(t) &=& \frac{1}{2i}\left( a_{p}^{k\ *}(t) a_{s}^{k}(t) - a_{s}^{k\ *}(t) a_{p}^{k}(t) \right) \\
S_{Z}^{k}(t) &=& \frac{1}{2}\left( a_{p}^{k\ *}(t) a_{p}^{k}(t) - a_{s}^{k\ *}(t) a_{s}^{k}(t) \right) \\
S_{0}^{k}(t) &=& \frac{1}{2}\left( a_{p}^{k\ *}(t) a_{p}^{k}(t) + a_{s}^{k\ *}(t) a_{s}^{k}(t) \right)
\end{eqnarray}
with $k$ being $r$ or $i$ for the reflected light and the incident light, respectively. Note that the Stokes parameter $S_{X}^{k}(t)$ measures the preponderance of the linearly polarized light $+\pi/4$-tilted from $Z$-axis (horizontal axis) over the one $-\pi/4$-tilted, while $S_{Z}^{k}(t)$ measures the preponderance of the $Z$-polarized light over the $X$-polarized one, and $S_{Y}^{k}(t)$ measures that of the right-hand circularly polarized one over the left-hand circularly polarized one. Here, and henceforth, we assume that $A(x,t)$, $B(x,t)$, and $C(x,t)$ are spatiotemporally-varying real parameters neglecting the imaginary parts of $r_{p}$ and $r_{s}$ (neglecting the absorption) for simplicity. As shown in Table~\ref{table:Optical_parameters} in  Sec.~\ref{sec:estimates}, the imaginary parts of $r_{p}$ and $r_{s}$ are indeed small for both insulators and metals when the incident angle $\theta_{1}$ is $\frac{\pi}{4}$. Equation~(\ref{eq:affine_map}) expresses an affine map when the Stokes parameters are viewed as the components of a 3-dimensional real vector. The reason why we use Stokes parameters based on classical field variables and avoid using the quantum-mechanical operators is that the affine map is non-unitary. In the case where the absorption is neglected and assuming $A(x,t)$, $B(x,t)$, and $C(x,t)$ being the real parameters, the non-unitary nature of the map Eq.~(\ref{eq:affine_map}) is the consequence of neglecting the refracted light at the boundary.

Now, let us assume that the incident light is $s$-polarized so that the expectation values of the Stokes parameters for the incident light are respectively given by
\begin{eqnarray}
\left\langle S_{X}^{i}(t) \right\rangle &=& 0 \label{eq:Sx0}\\
\left\langle S_{Y}^{i}(t) \right\rangle &=& 0 \label{eq:Sy0}\\
\left\langle S_{Z}^{i}(t) \right\rangle &=& -\frac{1}{2} \left| \beta_{s} \right|^{2} \label{eq:Sz0}\\
\left\langle S_{0}^{i}(t) \right\rangle &=& \frac{1}{2} \left| \beta_{s} \right|^{2} \label{eq:S00}
\end{eqnarray}
where the total incident photon flux is defined as $|\beta_{s}|^2=\frac{P_0}{\hbar\omega_0}$ with $P_0$ being the power of the incident $s$-polarized light. We then have the following expectation values of the Stokes parameters for the reflected light:
\begin{eqnarray}
\left[
\begin{array}{c}
\left\langle S_{X}^{r}(t) \right\rangle \\
\left\langle S_{Y}^{r}(t) \right\rangle \\
\left\langle S_{Z}^{r}(t) \right\rangle
\end{array}
\right] &=& \left[
\begin{array}{c}
B(x,t) C(x,t) \left| \beta_{s}  \right|^{2} \\
0 \\
\frac{C(x,t)^{2} - B(x,t)^{2}}{2} \left| \beta_{s}  \right|^{2} \\
\end{array}\right] \nonumber \\
&\sim& \left[
\begin{array}{c}
-\sqrt{2} \left( r_{p}-r_{s}\right)r_{s} \theta_{n}(x,t) \left| \beta_{s}  \right|^{2} \\
0 \\
-\frac{1}{2} r_{s}^{2} \left| \beta_{s}  \right|^{2} \\
\end{array}\right] = -\frac{1}{2} r_{s}^{2} \left| \beta_{s}  \right|^{2}\left[
\begin{array}{c}
2 \alpha(x,t) \\
0 \\
1 \\
\end{array}\right], \label{eq:affine_map_2}
\end{eqnarray}
where in the second line we use the fact that $\theta_{n}(x,t) \ll 1$ and retain the terms up to first order in $\theta_{n}(x,t)$. Here, $\alpha(x,t)$ is the polarization rotation angle defined in Eq.~(\ref{eq:rotation_angle}). This expression is the Stokes parameter analogue of Eq.~(\ref{eq:Fresnel2}). Now it is evident that measuring $\langle S_{X}^{r}(t) \rangle$ corresponds to measuring $\alpha(x,t)$.

Let us now examine the insensitivity of the polarization modulation when observing the path modulation in the setting (ii) in Fig.~1(a). Measuring with the setting amounts to measuring $\langle S_{Z}^{r}(t) \rangle$. Now, referring to Eq.~(\ref{eq:affine_map_2}), we can see that $\langle S_{Z}^{r}(t) \rangle$ is constant up to the first order in $\theta_{n}(t)$, showing the insensitivity to the polarization modulation in this measurement.

\begin{figure*}[t]
\begin{center}
\includegraphics[width=12cm,angle=0]{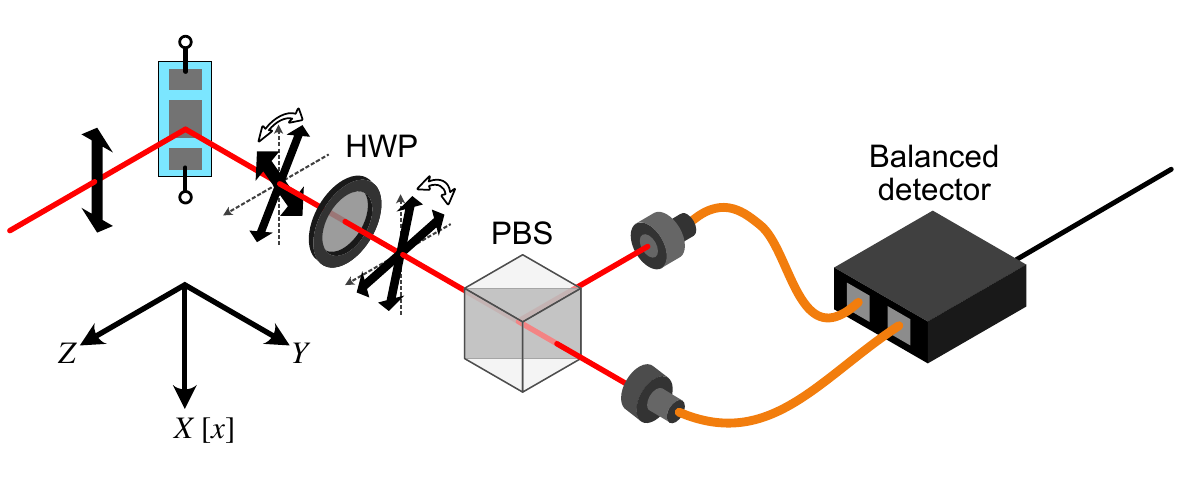}
\caption{Measurement setting (i) in Fig.~1(a). The coordinate system are denoted by $X$, $Y$, and $Z$ for the optical setup, where $X$ is equivalent to $x$ in Fig. 1.}
\label{fig:setup_zoom}
\end{center}
\end{figure*}

\subsection{Path modulation} \label{sec:path_mod}
When observing the path modulation in the setting (ii), what we actually measure is the intensity modulation of the light. Figures~\ref{fig:path_mod}(a) and \ref{fig:path_mod}(b) show how the path modulations of the reflected light, that is, $L(x,t)$ in Eq.~(\ref{eq:dl}), are converted into the intensity modulations after coupling the light into multi-mode fibers.

A coupler package (Thorlabs: F240APC) focuses the light collimated by the objective lens onto the end face of a multi-mode fiber (Thorlabs: M43L01). When the light is aligned properly as shown in Fig.~\ref{fig:path_mod}(a), all of the light is coupled into some propagation modes that are supported by the multi-fiber irrespective of the angle of incidence. On the other hand, when the light is deliberately misaligned as shown in Fig.~\ref{fig:path_mod}(b), the portion of light coupled into the fiber is sensitively depends on the angle of incidence which is proportional to the path modulation $ L(x,t)$.

The average optical power coupled into the fiber as a function of the $X$-position of the center of the coupler package is shown in the top panel of Fig.~\ref{fig:path_mod}(c).
The result is consistent with a specified value of the $X$-direction shift tolerance, $2\times f\times\tan(\rm{NA})=3.6$~mm, obtained using the numerical aperture of the multi-mode fiber $\rm{NA}=0.22$ and the focal length $f=7.93$ mm of the coupler package. The slope of the average optical power is maximum around $X=\pm 1.7$~mm, where the path modulation $L(x,t)$ is most effectively converted into the intensity modulation.

The middle and bottom panels of Fig.~\ref{fig:path_mod}(c) show respectively the amplitude and the phase of the intensity modulation signal at the SAW frequency as a function of the coupler-package position $X$. It can be seen that the intensity modulation occurs around the two positions, $X=\pm 1.7$~mm, where the slope of the average power shown in the top panel of Fig.~\ref{fig:path_mod}(c) are maximum. The phases of the two amplitude peaks are different by $\pi$ from each other, reflecting the fact that the corresponding slopes shown in the top panel of Fig.~\ref{fig:path_mod}(c) have opposite signs. Note that the setup of the path modulation measurement, that is, measurement in the setting (ii) in Fig.~1(a), was performed at $X \sim 1.9$~mm, which is close to the one of the optimal position $X \sim 1.7$~mm.

\begin{figure*}[t]
\begin{center}
\includegraphics[width=15cm,angle=0]{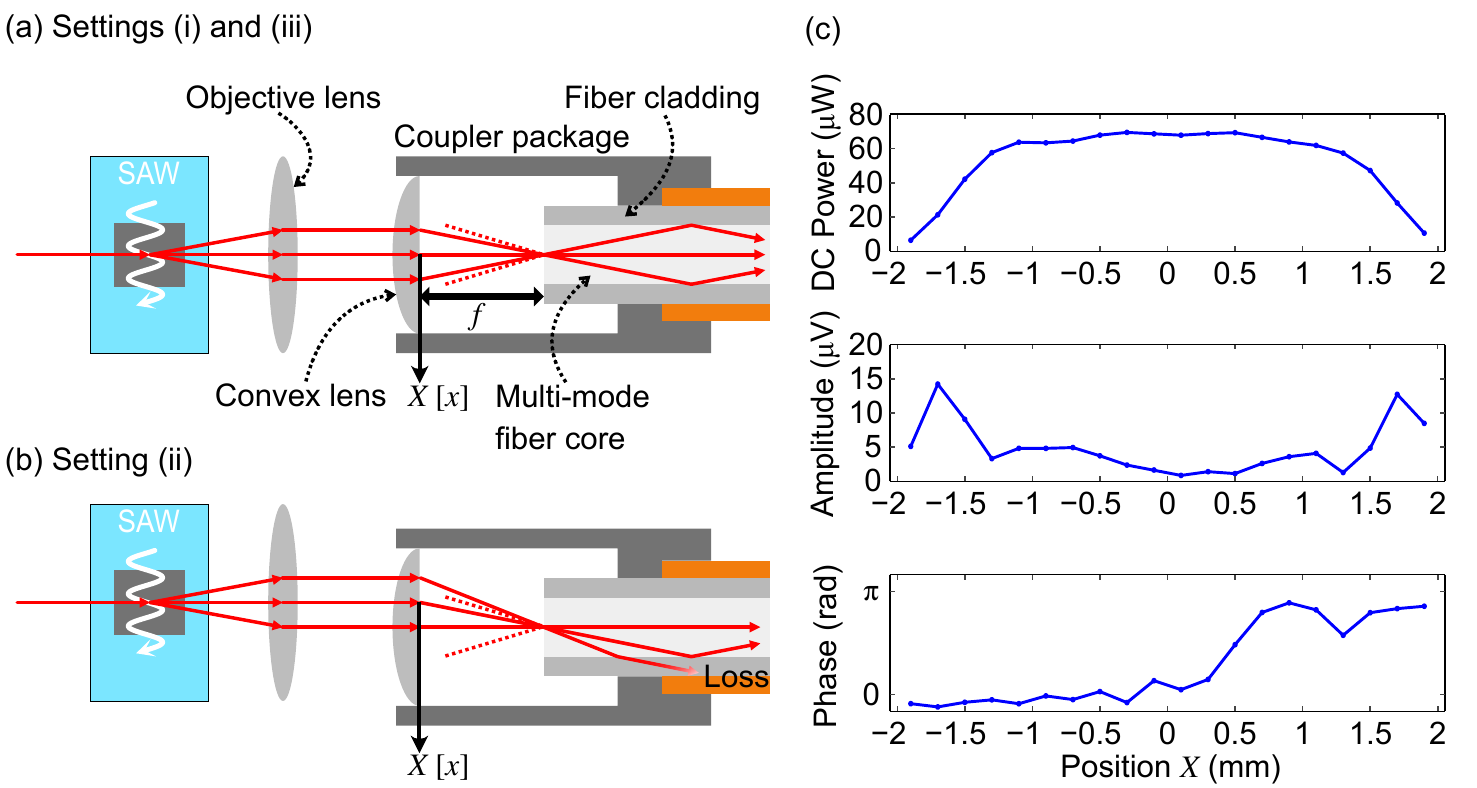}
\caption{(a) Details of the measurement settings (i) and (iii) in Fig.~1(a). (b) Details of the measurement setting (ii) in Fig.~1(a). The coordinate denoted by $X$ [$x$] for the optical setup is same as the ones used in Figs.~\ref{fig:S1} and~\ref{fig:setup_zoom} [Figs.~1 and 4].
The path of the reflected light varies in the $X$-axis, as shown in Fig.~4(a) in the main text. Three different wavevectors from the sample are depicted by red arrows. The light from the sample collimated by the objective lens is fed into the coupler package consisting of a convex lens with focal length $f$ and a multi-mode fiber. The red dashed lines indicate the angular range within which the light can be properly transmitted through the multi-mode fiber. Top panel of (c) shows the average optical power as a function of the position of the center of the coupler package, $X$. Middle and bottom panel of (c) show the amplitude and phase of the path modulation signal at the frequency $\omega_{\mathrm{SAW}}$.}
\label{fig:path_mod}
\end{center}
\end{figure*}

\subsection{Short summary} \label{sec:sum_mod}
We can conclude that, in the setting (i) in Fig.~1(a), what we measure is the Stokes parameter $\langle S_{X}^{r}(t) \rangle$, which, through Eq.~(\ref{eq:affine_map_2}), is proportional to the rotation angle of the polarization, with negligible path modulation because the light-to-fiber alignments are properly executed as illustrated in Fig.~\ref{fig:path_mod}(a). On the other hand, in the setting (ii), what we measure is the intensity modulation caused by the path modulation, which becomes only noticeable when deliberately misaligning the light into the fiber as in Fig.~\ref{fig:path_mod}(b) with negligible polarization modulation.

\section{SAW \lowercase{direction dependence}}\label{sec:saw_direction_dependence}

As shown in Fig.~\ref{fig:mapping}, the phase $\varphi$ linearly develops as a function of the probe position $x$, reflecting the fact that $\varphi(x) = q x$, where $q$ is the wave number of the SAW. Figures~\ref{fig:SM_Direction}(a) and (b) show similar spectra as in Fig.~\ref{fig:mapping}, but with the SAW propagating in the opposite direction (the SAW is excited through the IDT of Port 2). The opposite phase evolutions are observed in Fig.~\ref{fig:SM_Direction} compared with those in Fig.~\ref{fig:mapping}.

\begin{figure*}[t]
\begin{center}
\includegraphics[width=17cm,angle=0]{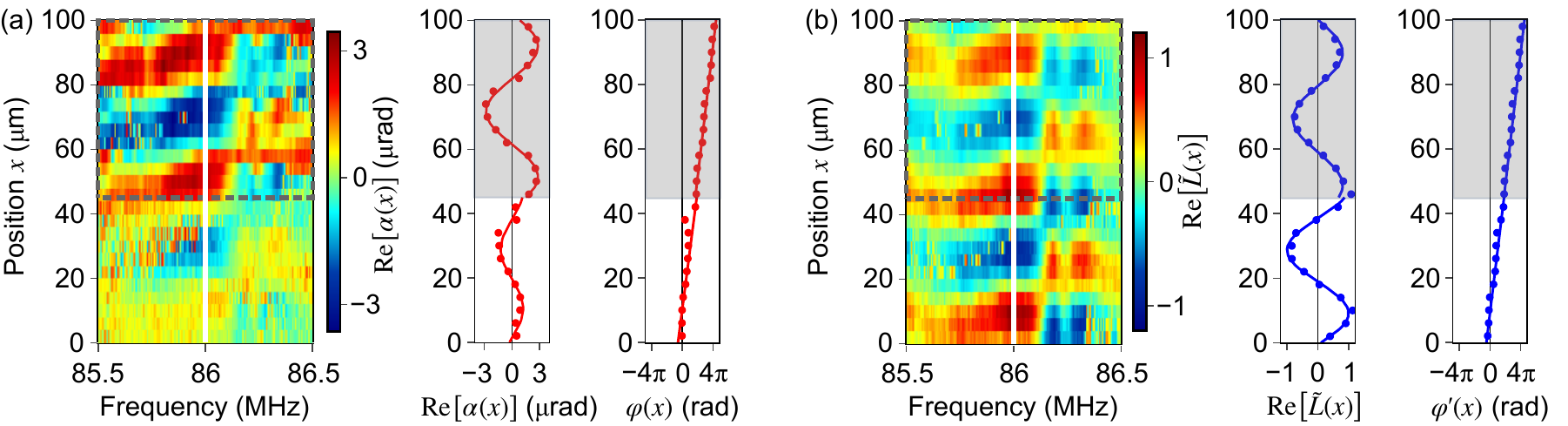}
\caption{
Position dependence of the SAW-induced optical modulation signal spectra observed through the polarimetric method (a) and the knife-edge method (b). (a)~Spectra of the polarization modulation $\mathrm{Re}\left[ \alpha(x) \right]$ acquired while the position of the beam spot $x$ is scanned across the boundary between the Al/$\linb$ region (enclosed by the dashed gray rectangle) and the bare $\linb$ region. (b)~Spectra of the normalized path modulation $\mathrm{Re}\left[ \tilde{L}(x) \right]$. Together with the color maps, cross-sections at the frequency of $\megahz{86}$ (indicated by white lines in the color maps) and their associated phases are shown with fitting curves. Note that here the SAW is excited through the IDT of Port 2. Otherwise, the measurement setup is identical to the one for Fig.~\ref{fig:mapping}.}
\label{fig:SM_Direction}
\end{center}
\end{figure*}

\section{S\lowercase{anity check of the models with experiments}}

\subsection{Dependence on the rf power of the SAW drive}\label{sec:saw_power_dependence}
From Eqs.~(\ref{eq:rotation_angle}) and (\ref{eq:dl}), we see that both amplitudes $\left|\alpha(x)\right|$ and $\left|L(x)\right|$ for the polarization modulation and for the path modulation linearly depend on the amplitude of the slope angle $\left|\theta_{n}(x)\right|$ and thus the amplitude of $\left|u_{z}(x)\right|$ through Eq.~(\ref{eq:theta_n2}). Since $\left|u_{z}(x)\right|$ is proportional to the square root of the driving rf power, we expect that both $\left|\alpha(x)\right|$ and $\left|L(x)\right|$ are proportional to the square root of the rf power. Figure~\ref{fig:SM_Power_saw} shows the relative amplitude $\left| \alpha(x) \right| / \left| \alpha_{\mathrm{ref}}(x) \right|$ for the polarization modulation (red dots) and $\left| L(x) \right|/ \left|L_{\mathrm{ref}}(x) \right|$ for the path modulation~(blue dots) as a function of the rf power. Here, $\left|\alpha_{\mathrm{ref}}(x)\right|$ and $\left|L_{\mathrm{ref}}(x)\right|$ denote the reference amplitudes with arbitrarily chosen rf powers. The linear fits on the log-log plots are also shown in Fig.~\ref{fig:SM_Power_saw}, indicating that the slopes are 0.51$\pm$0.02 for $\left| \alpha(x) \right| / \left| \alpha_{\mathrm{ref}}(x) \right|$~(red line) and 0.55$\pm$0.02 for $\left| L(x) \right|/ \left|L_{\mathrm{ref}}(x) \right|$~(blue line). The results validate the models given by Eqs.~(\ref{eq:rotation_angle}) and (\ref{eq:dl}), supporting that $\left|\alpha(x)\right|$ and $\left|L(x)\right|$ are proportional to  $\left|\theta_{n}(x)\right|$.

\begin{figure*}[t]
\begin{center}
\includegraphics[width=12cm,angle=0]{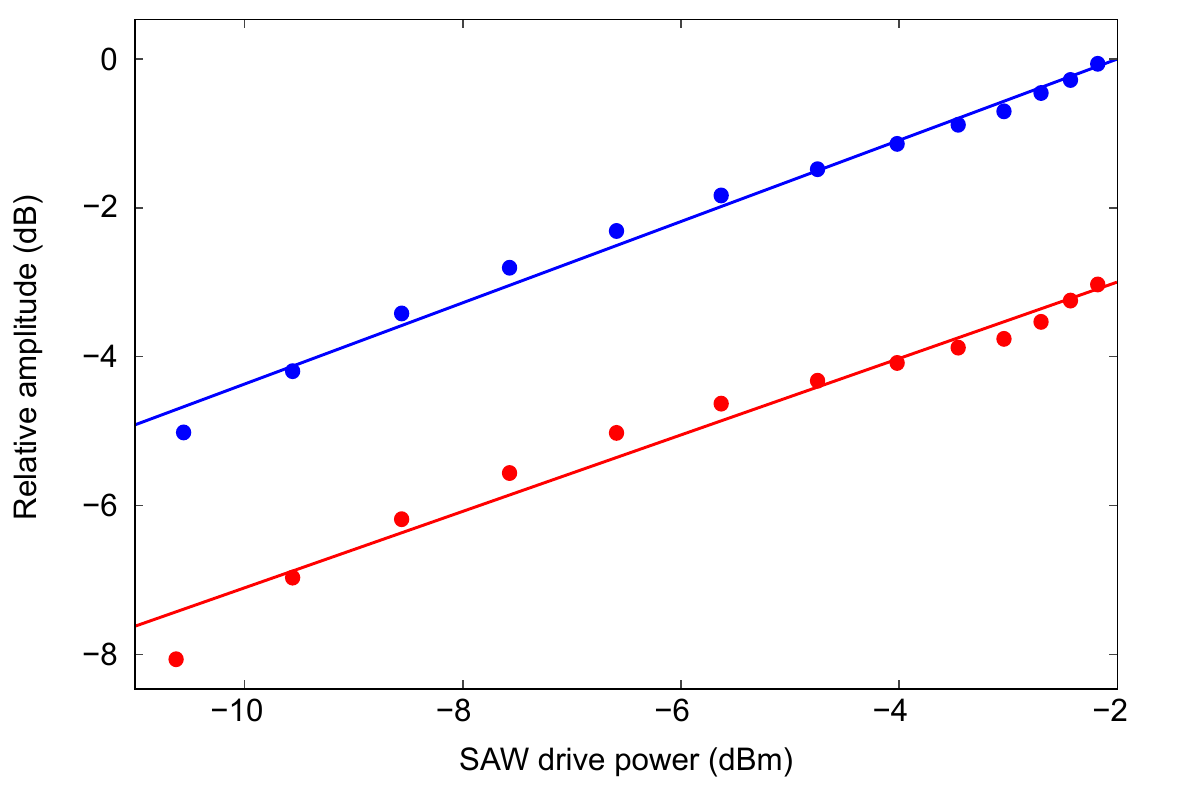}
\caption{Relative amplitudes for the polarization modulation (red dots) and the path modulation (blue dots) as a function of the rf power that is used for driving the SAW. The red and blue lines are the respective linear fits. The slopes of the fitted lines are 0.51$\pm$0.02 for the polarization modulation (red line) and 0.55$\pm$0.02 for the path modulation (blue line).}
\label{fig:SM_Power_saw}
\end{center}
\end{figure*}

\begin{figure*}[t]
\begin{center}
\includegraphics[width=12cm,angle=0]{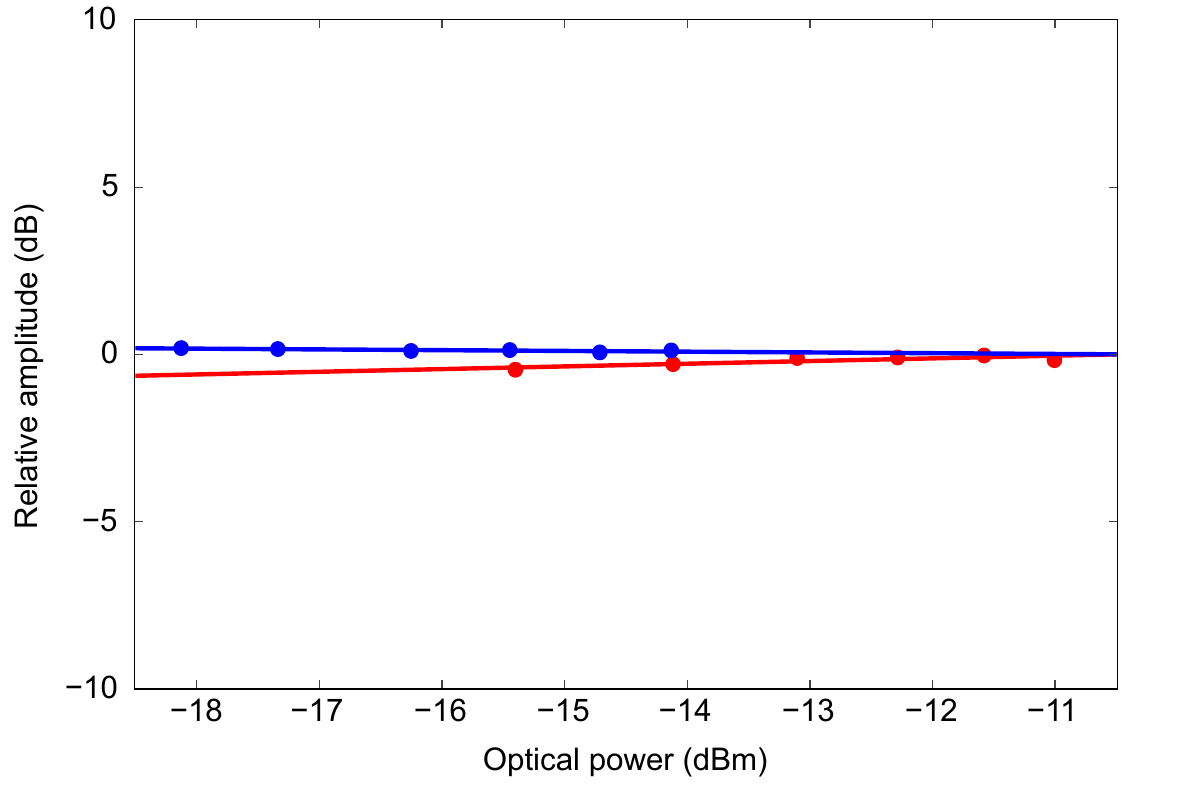}
\caption{Relative amplitudes for the polarization modulation (red dots) and the path modulation (blue dos) as a function of the detected optical power. The red and blue lines are the respective linear fits. The slopes of the fitted lines are $0.08\pm0.03$ for the polarization modulation (red line) and $-0.02\pm0.01$ for the path modulation (blue line). The range of the detected optical power in the two measurements varies by $3$~dB because the input power to the sample is set to be constant.}
\label{fig:SM_Power_light}
\end{center}
\end{figure*}

\subsection{Dependence on the optical power}\label{sec:optical_power_dependence}
We further check whether the amplitudes $\left|\alpha(x)\right|$ and $\left|L(x)\right|$ depend on the power of the incident light. Figure~\ref{fig:SM_Power_light} shows 
relative amplitude $\left|\alpha(x)\right|/ \left| \alpha_{\mathrm{ref}}'(x) \right|$ for the polarization modulation (red dots) and $\left|L(x)\right|/ \left|L_{\mathrm{ref}}'(x)\right|$ for the path modulation (blue dots) as a function of the detected optical power (which is proportional to the power of the incident light). Here, $\left|\alpha_{\mathrm{ref}}'(x)\right|$ and $\left|L_{\mathrm{ref}}'(x)\right|$ denote the reference amplitudes with arbitrarily chosen incident light powers. The linear fits on the log-log plots are also shown in Fig.~\ref{fig:SM_Power_light}, indicating that the slopes are $0.08\pm0.03$ for $\left|\alpha(x)\right|/ \left| \alpha_{\mathrm{ref}}'(x) \right|$~(red line) and $-0.02\pm0.01$ for $\left|L(x)\right|/ \left|L_{\mathrm{ref}}'(x)\right|$~(blue line). The results show the negligible non-linear dependence of the modulation amplitudes on the power of the incident light.

\section{C\lowercase{alibration of the measured modulation signals}}

\subsection{Calibration of the polarization modulation signal}\label{sec:pol_calibration}
Here, we shall discuss how the amount of polarization rotation can be calibrated using the shot noise level of the polarimetry as a reference. Note that the calibration scheme presented here is generic and can be used for other polarimetric measurements, e.g., for probing magneto-optical Kerr effect and Faraday effect.

As we discussed in Sec.\ref{sec:pol_mod}, what we actually measure in the setting (i) in Fig.~\ref{fig:setup}(a) is the instantaneous output voltage of the balanced detector. Here, the output voltage is proportional to the expectation value of the Stokes parameter $\langle S_{X}^{r}(t) \rangle$, that is,
\begin{equation}
V_{D}(t) \propto \langle S_{X}^{r}(t) \rangle.\label{eq:VD}
\end{equation}
For the shot-noise-based calibration scheme we need to deal with not just the expectation value of the the Stokes parameter $\langle S_{X}^{r}(t) \rangle$ but also the fluctuations including the shot-noise (vacuum-noise) contribution. To this end, let the dimensionless electric fields in Eq.~(\ref{eq:Fresnel}) be promoted to the quantum mechanical annihilation operators by the following simple prescription:
\begin{equation}
\left[ 
\begin{array}{c}
a_{p}^{i}(t) \\
a_{s}^{i}(t) 
\end{array}
\right]
\rightarrow
\left[ 
\begin{array}{c}
\hat{a}_{p}^{i}(t) \\
\hat{a}_{s}^{i}(t) 
\end{array}
\right] \label{eq:c2q_i}
\end{equation}
where we assume that the time-domain annihilation operator $\hat{a}_{j}^{i}(t)$ for the $j$-polarized light ($j=p, s$) satisfies the commutation relation, $[\hat{a}_{j}^{i}(t), \hat{a}_{j'}^{i}(t')]=\delta_{jj'}\delta(t-t')$. On the other hand, for the reflected light the similar prescription does not hold because the transformation from $a_{j}^{i}(t)$ to $a_{j}^{r}(t)$ given by Eq.~(\ref{eq:Fresnel}) is not unitary. To remedy this, let us introduce extra vacuum modes and modify Eq.~(\ref{eq:Fresnel}) so as to properly promote $a_{j}^{r}(t)$ into the quantum mechanical annihilation operators $\hat{a}_{j}^{r}(t)$:
\begin{equation}
\left[
\begin{array}{c}
a_{p}^{r}(t)  \\
a_{s}^{r}(t) 
\end{array}
\right] \rightarrow
\left[
\begin{array}{c}
\hat{a}_{p}^{r}(t)  \\
\hat{a}_{s}^{r}(t) 
\end{array}
\right] = \left[
\begin{array}{cc}
A(x,t) & C(x,t) \\
C(x,t) & B(x,t)
\end{array}
\right]\left[
\begin{array}{c}
\hat{a}_{p}^{i}(t)  \\
\hat{a}_{s}^{i}(t) 
\end{array}
\right] + \left[
\begin{array}{c}
D(x,t) \hat{b}_{p}(t)  \\
E(x,t) \hat{b}_{s}(t) 
\end{array}
\right],\label{eq:c2q_r}
\end{equation}
where 
\begin{eqnarray}
D(x,t) &=& \sqrt{1-A(x,t)^{2} - C(x,t)^{2}} \\
E(x,t) &=& \sqrt{1-B(x,t)^{2} - C(x,t)^{2}}.
\end{eqnarray}
Here, the annihilation operators $\hat{b}_{p}(t)$ and $\hat{b}_{s}(t)$ are for the extra vacuum modes, which are introduced to make the operators $\hat{a}_{j}^{r}(t)$ satisfy the usual commutation relation, that is, $[\hat{a}_{j}^{r}(t), \hat{a}_{j'}^{r}(t')]=\delta_{jj'}\delta(t-t')$. 
The Stokes parameters $S_{j}^{k}(t)$ are then accordingly modified into Stokes operators $\hat{S}_{j}^{k}(t)$ with the prescriptions Eqs.~(\ref{eq:c2q_i}) and (\ref{eq:c2q_r}). Note that these Stokes parameters satisfy the commutation relations $[\hat{S}_{i}^{k}(t), \hat{S}_{j}^{k}(t')]= i \epsilon_{ijl} \hat{S}_{l}^{k}(t) \delta(t-t')$, where $\epsilon_{ijl}$ is the Levi-Civita symbol. The Stokes operator $\hat{S}_{X}^{r}(t)$ in Eq.~(\ref{eq:VD}) can then be expressed as
\begin{eqnarray}
\hat{S}_{X}^{r}(t) &=& \frac{1}{2} \left(\hat{a}_{p}^{r \ \dagger}(t) \hat{a}_{s}^{r}(t) +\hat{a}_{s}^{r\ \dagger}(t) \hat{a}_{p}^{r}(t) \right) \nonumber \\
&=& \frac{1}{2} \left(\left( A\hat{a}_{p}^{i \ \dagger} + C \hat{a}_{s}^{i \ \dagger} + D \hat{b}_{p}^{\dagger}\right) \left(C \hat{a}_{p}^{i} +B \hat{a}_{s}^{i} + E \hat{b}_{s} \right)  +\left(C\hat{a}_{p}^{i\ \dagger} + B\hat{a}_{s}^{i\ \dagger} + E \hat{b}_{s}^{\dagger} \right) \left(A \hat{a}_{p}^{i}+ C \hat{a}_{s}^{i} + D \hat{b}_{p} \right) \right) \nonumber \\
&=& \left( AB+C^{2}\right)\hat{S}_{X}^{i}(t) + C \left( A-B\right) \hat{S}_{Z}^{i}(t) + C \left( A+B \right) \hat{S}_{0}^{i}(t) \nonumber \\
&\ & \ \ \ + \frac{1}{2} BD \left(\hat{a}_{s}^{i\ \dagger}(t) \hat{b}_{p}(t) + \mathrm{h.c.} \right) + \frac{1}{2} CE \left( \hat{a}_{s}^{i\ \dagger}(t) \hat{b}_{s}(t) + \mathrm{h.c.} \right) \nonumber \\
&\ & \ \ \ +\frac{1}{2} CD \left(\hat{a}_{p}^{i\ \dagger}(t) \hat{b}_{p}(t) + \mathrm{h.c.} \right) + \frac{1}{2} AE \left( \hat{a}_{p}^{i\ \dagger}(t) \hat{b}_{s}(t) + \mathrm{h.c.} \right) + \frac{1}{2} DE \left(\hat{b}_{s}^{\dagger}(t) \hat{b}_{p}(t)  + \mathrm{h.c.} \right). \label{eq:Sx}
\end{eqnarray}

Assuming again that the incident light is $s$-polarized, the annihilation operator $\hat{a}_{s}^{i}(t)$ can then be split into the classical part $\beta_{s}$ and the part representing the quantum fluctuation $\hat{d}_{s}(t)$ as
\begin{equation}
\hat{a}_{s}^{i}(t) = \beta_{s} + \hat{d}_{s}(t),
\end{equation}
where $\beta_{s}$ is related to the power $P_0$ of the incident $s$-polarized light as $|\beta_{s}|^2=\frac{P_0}{\hbar\omega_0}$. Henceforth, we shall assume the classical part $\beta_{s}$ to be real for simplicity. As for the other annihilation operators we note that there are only parts representing quantum fluctuations. Suppressing higher order terms in $\theta_{n}(x,t)$ and terms that contain quantum fluctuations only, the Stokes operator, Eq~(\ref{eq:Sx}), can be simplified as
\begin{eqnarray}
\hat{S}_{X}^{r}(t) &\sim& r_{p}r_{s}\underbrace{\hat{S}_{X}^{i}(t)}_{\frac{1}{2}\beta_{s}\left(\hat{a}_{p}^{i}(t) +\hat{a}_{p}^{i \dagger}(t) \right)} - \sqrt{2}\left( r_{p}-r_{s}\right)^{2}\theta_{n}(x,t) \underbrace{\hat{S}_{Z}^{i}(t)}_{-\frac{1}{2}\beta_{s}^{2}} - \sqrt{2} \left( r_{p}+r_{s} \right)\left( r_{p}-r_{s}\right)\theta_{n}(x,t) \underbrace{\hat{S}_{0}^{i}(t)}_{\frac{1}{2}\beta_{s}^{2}} \nonumber \\
&\ & \ \ \ + r_{s}\sqrt{1-r_{p}^{2}}\left(\frac{1}{2}\beta_{s} \left(\hat{b}_{p}(t) + \hat{b}_{p}^{\dagger}(t) \right) \right). \label{eq:Sx2}
\end{eqnarray}
Based on Eqs.~(\ref{eq:VD}) and (\ref{eq:Sx2}) the auto-correlation of $\hat{V}_{D}(t)$ can be given by
\begin{eqnarray}
\langle \hat{V}_{D}(t) \hat{V}_{D}(t+\tau)\rangle
&\propto&
\langle \hat{S}_{X}^{r}(t) \hat{S}_{X}^{r}(t+\tau)\rangle  \nonumber \\
&\sim& \frac{1}{4} r_{p}^{2}r_{s}^{2} \beta_{s}^2 \left( \underbrace{\langle \hat{a}_{p}^{i}(t) \hat{a}_{p}^{i}(t+\tau) \rangle}_{0} + \underbrace{\langle \hat{a}_{p}^{i}(t) \hat{a}_{p}^{i \dagger}(t+\tau) \rangle}_{\delta (\tau)} + \underbrace{\langle \hat{a}_{p}^{i \dagger}(t) \hat{a}_{p}^{i}(t+\tau) \rangle}_{0} + \underbrace{\langle \hat{a}_{p}^{i \dagger}(t) \hat{a}_{p}^{i \dagger}(t+\tau) \rangle}_{0} \right) \nonumber \\ 
&\ & \ \ \ + 2\left( r_{p}-r_{s} \right)^{2}r_{s}^{2}\beta_{s}^4 \underbrace{\langle \theta_{n}(x,t) \theta_{n}(x,t+\tau) \rangle}_{\frac{1}{2} \left|\theta_{n}(x)\right|^{2} \cos \left( \omega_{\mathrm{SAW}}\tau \right)} \nonumber \\
&\ & \ \ \ + \frac{1}{4} \left( 1-r_{p}^{2} \right) r_{s}^{2} \beta_{s}^2 \left( \underbrace{\langle \hat{b}_{p}(t) \hat{b}_{p}(t+\tau) \rangle}_{0} + \underbrace{\langle \hat{b}_{p}(t) \hat{b}_{p}^{\dagger}(t+\tau) \rangle}_{\delta(\tau)} + \underbrace{\langle \hat{b}_{p}^{\dagger}(t) \hat{b}_{p}(t+\tau) \rangle}_{0} + \underbrace{\langle \hat{b}_{p}^{\dagger}(t) \hat{b}_{p}^{\dagger}(t+\tau) \rangle}_{0} \right), \label{eq:VDVD}
\end{eqnarray}
where the auto-correlation of the operators for the light are evaluated with respect to the vacuum states. As for the auto-correlation of the slope angle $\theta_{n}(x,t)$ we took the form given by Eqs.~(\ref{eq:theta_n2}) and (\ref{eq:theta_n3}) and evaluated its auto-correlation. Consequently, we have
\begin{equation}
\langle \hat{V}_{D}(t) \hat{V}_{D}(t+\tau)\rangle
\propto \frac{1}{4}r_{s}^{2} \beta_{s}^2\delta(\tau) +\frac{1}{2}(r_{p}-r_{s})^{2}r_{s}^{2} \beta_{s}^4 \left| \theta_{n}(x)\right|^2 (e^{i\omega_{\rm{SAW}} \tau}+e^{-i\omega_{\rm{SAW}} \tau}).
\end{equation}
Fourier transforming $\langle \hat{V}_{D}(t) \hat{V}_{D}(t+\tau)\rangle$, we obtain the following power spectrum $S_{VV}(\omega)$:
\begin{eqnarray}
S_{VV}(\omega)\propto\frac{1}{4}r_{s}^{2}\beta_{s}^2 + \frac{1}{2}\left( r_{p}-r_{s}\right)^{2} r_{s}^{2}\left|\theta_{n}(x)\right|^2 \beta_{s}^4 \left[ 2\pi \delta \left(\omega - \omega_{\rm{SAW}} \right) + 2\pi \delta \left(\omega + \omega_{\rm{SAW}} \right) \right].
\end{eqnarray}
At resonance $\omega = \omega_{\rm{SAW}}$ the spectral power within the bandwidth $\Delta f \equiv \frac{\Delta \omega}{2\pi}$ reads,\cite{Hisatomi2016}
\begin{equation}
S_{VV}(\omega_{\rm{SAW}})\Delta f \propto \underbrace{\frac{1}{4}r_{s}^{2}\beta_{s}^2 \Delta f}_{\mathrm{noise}} +\underbrace{\frac{1}{2}\left(r_{p}-r_{s} \right)^{2}r_{s}^{2} \left|\theta_{n}(x)\right|^{2} \beta_{s}^4}_{\mathrm{signal}},
\end{equation}
where the first term is the frequency-independent shot noise and the second term is the signal due to the coherent SAW excitation. Since from Eq.~(\ref{eq:rotation_angle}) the amplitude of the slope angle $\theta_{n}(x)$ can be expressed in term of the amplitude of polarization rotation $\alpha(x)$ as
\begin{equation}
\theta_{n}(x) = \frac{1}{\sqrt{2}} \frac{r_{s}}{r_{s}-r_{p}}\alpha(x) \label{eq:alpha_x}
\end{equation}
the spectral power can also be given by
\begin{equation}
S_{VV}(\omega_{\rm{SAW}})\Delta f \propto \underbrace{\frac{1}{4}r_{s}^{2}\beta_{s}^2 \Delta f}_{\mathrm{noise}} +\underbrace{\frac{1}{4}r_{s}^{4}\left|\alpha(x)\right|^{2} \beta_{s}^4}_{\mathrm{signal}}.
\end{equation}
Thus, we have the following expression of the signal-to-noise ratio (SNR):
\begin{equation}
\mathrm{SNR} = \frac{ r_{s}^{2}\beta_{s}^2 \left|\alpha(x)\right|^{2}}{\Delta f}. \label{eq:SNR}
\end{equation}
Thus, by experimentally evaluating SNR we can evaluate the amplitude of the rotation angle $\alpha(x)$ with the known value of the photon flux $\beta_{s}^2$ and the measurement bandwidth $\Delta f$.

Note that in reality there are inevitable losses of photons other than the loss in the reflection at the surface of the device. These losses include the ones in collecting reflected light through the objective lens and in coupling the light into the multi-mode fibers. As far as the losses incur similarly in $p$- and $s$-polarized light, they are effectively incorporated by a single amplitude loss factor $\eta$ in Eq.~(\ref{eq:Fresnel}) as
\begin{equation}
\left[
\begin{array}{c}
a_{p}^{r}(t)  \\
a_{s}^{r}(t) 
\end{array}
\right] = R(-\sqrt{2}\theta_n(t))\left[
\begin{array}{cc}
\eta r_p & 0 \\
0 & \eta r_s \\
\end{array}
\right] 
R(\sqrt{2}\theta_n(t))
\left[
\begin{array}{c}
a_{p}^{i}(t)  \\
a_{s}^{i}(t) 
\end{array}
\right].\label{eq:Fresnel3}
\end{equation}
All the procedures to calibrate the rotation angle $\alpha(x)$ we have discussed in this section remain intact but replacing $r_{p} \rightarrow \eta r_{p}$ and $r_{s} \rightarrow \eta r_{s}$. In particular, $\alpha(x)$ defined by Eq.~(\ref{eq:alpha_x}) is obviously insensitive to these losses. In calibrating $\alpha(x)$ we thus employ the modified form of the SNR expression:
\begin{equation}
\mathrm{SNR} = \frac{ \eta^{2} r_{s}^{2}\beta_{s}^2 \left|\alpha(x)\right|^{2}}{\Delta f}, \label{eq:SNR2}
\end{equation}
where we evaluate the SNR with a separately measured value of the photon flux of the reflected light \textit{at the balanced detector} $\beta^{2} = \eta^{2} r_{s}^{2}\beta_{s}^2$, which incorporates all the losses, and the measurement bandwidth $\Delta f$.

\begin{figure*}[b]
\begin{center}
\includegraphics[width=12cm,angle=0]{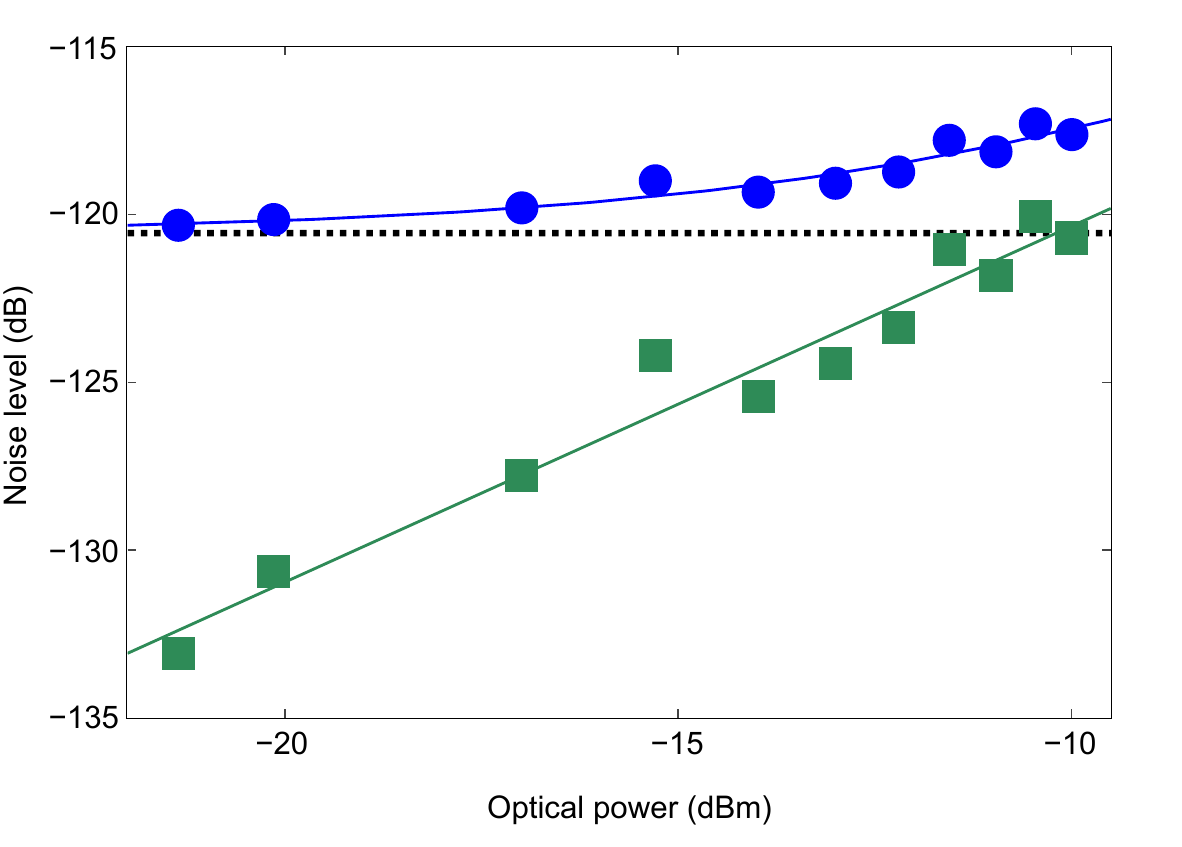}
\caption{Total noise level (blue circles) and shot noise level (green squares) as a function of the optical power at the measurement frequency $\omega/2\pi = 86$~MHz within the bandwidth $\Delta \omega/2\pi = 10$~Hz. The shot noise level is obtained by subtracting the separately measured electronic noise (dotted line). The shot noise level grows linearly with the laser power as indicated by the green line (the slope of the line is 1.06$\pm$0.08). The total noise level can be fitted by the sum of the fitted shot noise level (green line) and the electronic noise (dotted line) as shown by the blue line.
}
\label{fig:shot-noise}
\end{center}
\end{figure*}

Figure~\ref{fig:shot-noise} shows the total noise level (blue circles) as a function of the optical power at the measurement rf frequency $\omega/2\pi= 86$~MHz within the bandwidth $\Delta f = 10$~Hz. The shot noise level and the electronic noise level are comparable. The shot noise level (green squares) is obtained by subtracting the electronic noise (dotted line). The left panel of Fig.~\ref{fig:SM_Reflection}(a) shows the optical power reflected off the device surface, collected by the objective lens, and detected by the detector, that is, $\hbar \omega_{0}\beta^{2} = \hbar \omega_{0} \eta^{2} r_{s}^{2} \beta_{s}^{2}$, as a function of the position of the beam spot, $x$. From these reflected powers we can assign a shot noise level for each position. We then compare the signal level as shown in the middle panel of Fig.~\ref{fig:SM_Reflection}(a) with the corresponding shot noise level and then evaluate the SNR. The signal in units of volts in the middle panel of Fig.~\ref{fig:SM_Reflection}(a) can then be calibrated in light of Eq.~(\ref{eq:SNR2}). The real part of the resultant calibrated signal
\begin{equation}
\mathrm{Re}\left[ \alpha(x) \right] = \left| \alpha (x) \right| \cos \left(\varphi (x) \right)
\end{equation}
in units of radians is shown in the right panel of Fig.~\ref{fig:SM_Reflection}(a) with the fitting curves $\alpha_{\mathrm{Al}}\cos(q_{\mathrm{Al}} x)$ for the Al/LiNbO$_{3}$ region and $\alpha_{\mathrm{LN}}\cos(q_{\mathrm{LN}} x)$ for the bare LiNbO$_{3}$ region, where $\alpha_{\mathrm{Al}}$, $\alpha_{\mathrm{LN}}$, $q_{\mathrm{Al}}\equiv \frac{2 \pi}{\lambda_{\mathrm{Al}}}$, and $q_{\mathrm{LN}}\equiv \frac{2 \pi}{\lambda_{\mathrm{LN}}}$ are the fitting parameters for the respective regions.

\begin{figure*}[b]
\begin{center}
\includegraphics[width=12cm,angle=0]{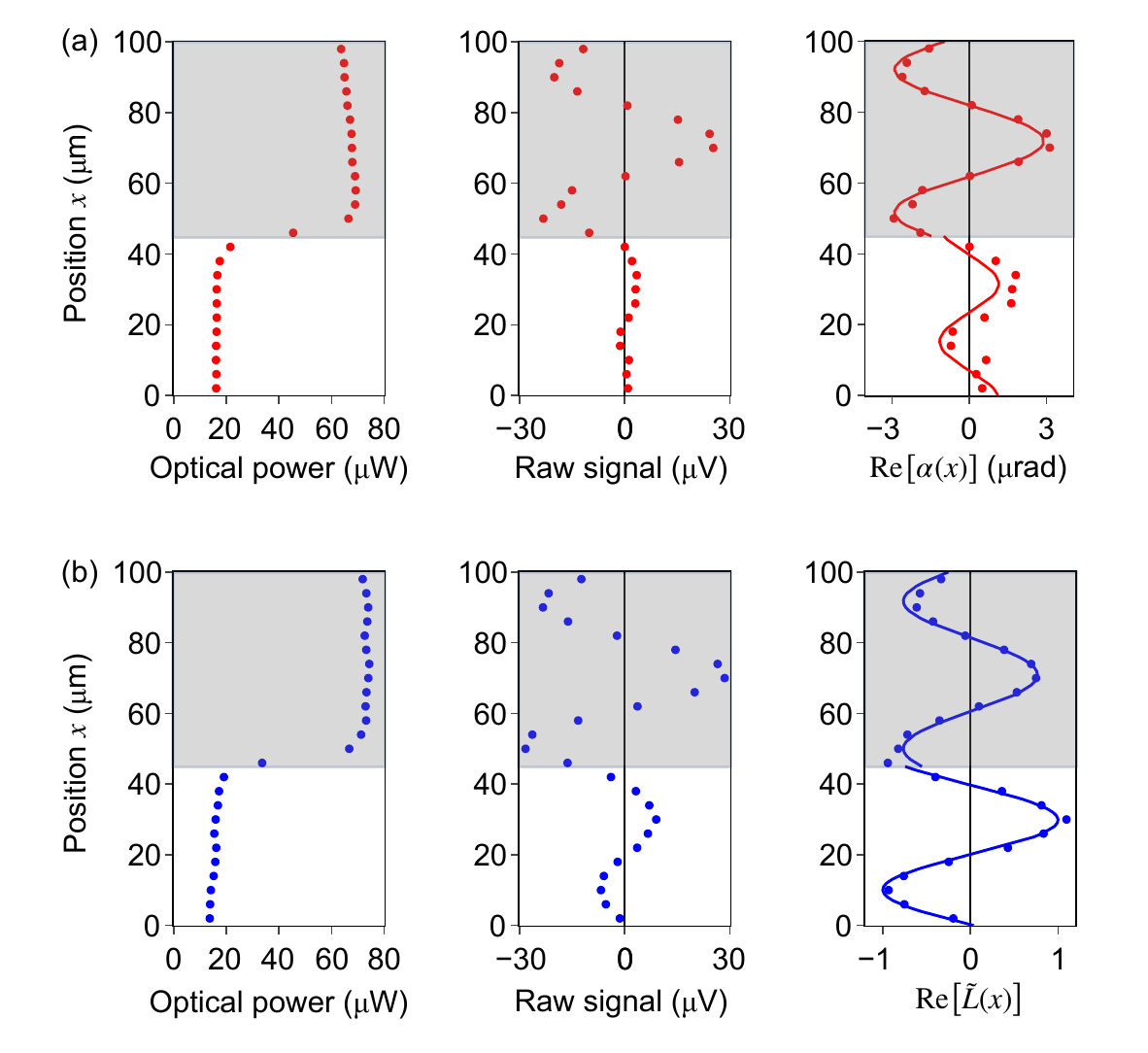}
\caption{(a) Calibration of the polarization modulation signal. Left panel shows the optical power reflected off the surface as a function of the position of the beam spot. The raw signal in units of $\mu$V and the calibrated signal $\mathrm{Re}\left[ \alpha(x) \right]$ with the fitting curve [the same signal as in Fig.~\ref{fig:mapping}(a)] in units of $\mu$rad are shown in the middle and the right panels, respectively. (b) Calibration of the path modulation signal. Left panel shows the optical power reflected off the surface as a function of the position of the beam spot. The raw signal in units of $\mu$V and the calibrated dimensionless signal $\mathrm{Re}\left[ \tilde{L}(x) \right]$ with the fitting curve [the same signal as in Fig.~\ref{fig:mapping}(b)] are shown in the middle and the right panels, respectively.}
\label{fig:SM_Reflection}
\end{center}
\end{figure*}

\subsection{Calibration of the path modulation signal}\label{sec:path_calibration}
The left panel of Fig.~\ref{fig:SM_Reflection}(b) shows the optical power reflected off the device surface, collected by the objective lens, and detected by the detector, that is, $\hbar \omega_{0}\beta^{2} = \hbar \omega_{0} \eta^{2} r_{s}^{2} \beta_{s}^{2}$, as a function of the position of the beam spot, $x$. The signal in units of volts in the middle panel of Fig.~\ref{fig:SM_Reflection}(b) is first divided by the detected optical power and then the maximum value $\left|L(x_{\mathrm{max}})\right|$ is normalized to unity. The real part of the resultant dimensionless signal
\begin{equation}
\mathrm{Re}\left[ \tilde{L}(x) \right] \equiv \mathrm{Re}\left[ \frac{L(x)}{\left|L(x_{\mathrm{max}} \right|} \right] = \frac{ \left|L(x)\right|}{ \left|L(x_{\mathrm{max}}) \right|} \cos \left(\varphi' (x) \right) = \left|\tilde{L}(x) \right| \cos \left(\varphi' (x) \right)
\end{equation}
is shown in the right panel of Fig.~\ref{fig:SM_Reflection}(b) with the fitting curves $\tilde{L}_{\mathrm{Al}}\cos(q'_{\mathrm{Al}} x)$ for the Al/LiNbO$_{3}$ region and $\tilde{L}_{\mathrm{LN}}\cos(q'_{\mathrm{LN}} x)$ for the bare LiNbO$_{3}$ region, where $\tilde{L}_{\mathrm{Al}}$, $\tilde{L}_{\mathrm{LN}}$, $q'_{\mathrm{Al}}\equiv \frac{2 \pi}{\lambda'_{\mathrm{Al}}}$, and $q'_{\mathrm{LN}}\equiv \frac{2 \pi}{\lambda'_{\mathrm{LN}}}$ are the fitting parameters for the respective regions.

\section{E\lowercase{stimate of the amplitude of the surface slope angle}}\label{sec:estimates}
In this section, to check the validity and reliability of our model of polarization modulation in Sec.~\ref{sec:polarization} and calibration method in Sec.~\ref{sec:pol_calibration}, we shall compare the surface slope angle $\theta_n(x,t)$ obtained from the polarization modulation experiment and that deduced from the physical dimension of the device and the SAW driving power.

From the experiments, the amplitudes of the polarization rotation angle
$\alpha_{\mathrm{Al}}= 2.91\pm 0.09$ ($2.82\pm 0.12$)~$\mu$rad and $\alpha_{\mathrm{LN}}= 1.15\pm 0.34$ ($1.22\pm 0.17$)~$\mu$rad with the wavelengths $\lambda_{\mathrm{Al}} = 40.3 \pm 0.6$ ($40.5 \pm 0.7$)~$\mu$m and $\lambda_{\mathrm{LN}} = 32.7 \pm 4.0 $ ($36.8 \pm 2.7$)~$\mu$m for SAW propagation in the $+x$ ($-x$)-direction are obtained, respectively. From Eqs.~(\ref{eq:rp_2}), (\ref{eq:rs_2}), and (\ref{eq:alpha_x}) and the material-dependent parameters listed in Table~\ref{table:Optical_parameters}, we then obtain the amplitudes of the surface slope angle
\begin{equation}
\theta_{n,\mathrm{Al}}= 1.04\pm 0.03\ (1.00\pm 0.04)~\mu\mathrm{rad} \label{eq:theta_al}
\end{equation}
for the spot position $x$ in the Al/LiNbO$_{3}$ region and
\begin{equation}
\theta_{n,\mathrm{LN}}= 0.54\pm 0.16\ (0.58\pm 0.08)~\mu\mathrm{rad} \label{eq:theta_ln}
\end{equation}
for the spot position $x$ in the bare LiNbO$_{3}$ region, respectively, when the SAW propagates in the $+x$ ($-x$)-direction. Note here that the imaginary parts of $r_{p}$ and $r_{s}$ are small even when the imaginary parts of the refractive indices is large (as for Al). Thus, we have neglected and will neglect the imaginary parts of $r_{p}$ and $r_{s}$ for simplicity.

\begin{table}[h]
  \begin{center}
    \caption{Optical parameters of Al and $\linb$} 
    \begin{tabular}{|c||c|c|} \hline
      Parameter & Al & $\linb$  \\ \hline \hline
      Refractive index & $1.58+ 15.7 i$ & $2.21$ \\ \hline 
      $r_p$ with $\theta_1=\pi/4$ & $0.967+0.174 i$ & $0.245$\\ \hline
      $r_s$ with $\theta_1=\pi/4$& $-(0.987+0.0883 i)$  & $-0.495$\\ \hline
    \end{tabular}
  \label{table:Optical_parameters}
  \end{center}
\end{table}

Now we compare these experimentally-obtained values with the values deduced from the physical dimensions of the SAW device with a given driving power. For simplicity we only consider the the surface slope $\theta_{n}$ incurred by the displacement of the LiNbO$_{3}$ substrate. Since the thickness of the Al film (100~nm) is far thinner compared with the SAW wavelength (40~$\mu$m), considering only the displacement of the bare LiNbO$_{3}$ would be warranted.

The amplitude of the displacement field $|u_z(x)|$ defined by Eq.~(\ref{eq:u_z}) can be obtained by identifying $|u_z(x)|$ as a displacement of a simple harmonic oscillator with the zero-point-fluctuation $U_{0}$, namely~\cite{SK2015}
\begin{equation}
|u_z(x)| = U_{0} \sqrt{N_{\rm{ph}}}. \label{eq:uz_abs}
\end{equation}
Here, $N_{\rm{ph}}$ is the number of phonons excited in the effective mode volume
\begin{equation}
V= \underbrace{w}_{\mathrm{width}} \times \underbrace{v_{\mathrm{SAW}}\ t}_{\mathrm{length}} \times \underbrace{\lambda_\mathrm{SAW}}_{\mathrm{depth}}, \label{eq:v}
\end{equation}
with $w$ being the width of the IDT, $\lambda_{\mathrm{SAW}}$ being the wavelength of the SAW, and $v_{\mathrm{SAW}}=3.5\,\rm{km/s}$ being the phase velocity of the SAW propagating along the crystalline $Z$-axis on the $\linb$ substrate. For the SAW device having the parameters listed in Table~\ref{table:Device_parameters}, we have $V=1.4\times 10^{-4}$~m$^{3}$ for $t=1$~s. Note that since only the SAW propagating in either of the $\pm x$-directions excited by the IDT is used, the power delivered into the SAW is $3\,\rm{dB}$ lower than the RF power applied to the IDT ($-10$ dBm). The energy stored in the mode volume $V$ in one second is then estimated to be $0.05\,\rm{mJ}$. Dividing this energy by $\hbar \omega_{\mathrm{SAW}}$, we have
$N_{\rm{ph}}\sim 8.8 \times10^{20}$. With the effective mode volume $V$ given by Eq.~(\ref{eq:v}), the zero-point-fluctuation $U_{0}$ can then be read as
\begin{equation}
U_{0} = \sqrt{\frac{\hbar}{2 \rho V \omega_{\mathrm{SAW}}}} \sim 3.9 \times 10^{-22}~\mathrm{m}
\end{equation}
where $\rho=4.65$~g/cm$^{3}$ is the mass density of the $\linb$. With this value of $U_{0}$, from Eq.~(\ref{eq:uz_abs}), we have
\begin{equation}
\left| u_{z}(x) \right| \sim 1.1 \times 10^{-11}~\mathrm{m}.
\end{equation}

\begin{table}[h]
  \begin{center}
    \caption{SAW device parameters.} 
    \begin{tabular}{|l|c|} \hline
      Parameter & Value \\ \hline \hline
      Frequency: $\omega_{\mathrm{SAW}}/2\pi$ & 86\,\rm{MHz}\\ 
      Wavelength: $\lambda$ & 40\,\rm{$\mathrm{\mu}$m} \\ 
      IDT width: $w$ & 1\,\rm{mm} \\ \hline
    \end{tabular}
  \label{table:Device_parameters}
  \end{center}
\end{table}

The amplitude of the displacement field in Eq,~(\ref{eq:u_z}) can now be written as 
\begin{equation}
u_z(x) = \left|u_{z}(x)\right|e^{iqx} = U_{0} \sqrt{N_{\mathrm{ph}}} e^{iqx}
\end{equation} 
where $q=\frac{2\pi}{\lambda}$ is the wavevector. The amplitude of the slope angle $\left|\theta_n(x)\right|$ in Eq.~(\ref{eq:theta_n3}) can then be obtained as
\begin{equation}
  \left|\theta_n(x)\right| = \left|\frac{\partial u_{z}(x)}{\partial x} \right| = q\,|u_z(x,t)|\sim 1.8~\mu\mathrm{rad}. \label{eq:theta_t}
\end{equation}

Given that the calibrated values of the slope angles $\theta_{n,\mathrm{Al}}$ in the Al/LiNbO$_3$ region and $\theta_{n,\mathrm{LN}}$ in the bare LiNbO$_3$ region in Eqs.~(\ref{eq:theta_al}) and (\ref{eq:theta_ln}), the resultant value in Eq.~(\ref{eq:theta_t}) is in the reasonable agreement, suggesting the validity of the model and the calibration method discussed in Sec.~\ref{sec:polarization} and \ref{sec:pol_calibration}. There is a discrepancy between $\theta_{n,\mathrm{Al}}$ and $\theta_{n,\mathrm{LN}}$, though. The discrepancy may be caused by the difference between the amplitudes of the displacement $\left|u_{z}(x) \right|$ in the Al/LiNbO$_{3}$ region and that in the LiNbO$_{3}$ region due to, e.g., extra reflection at the boundary between the Al/LiNbO$_{3}$ region and the bare LiNbO$_3$ region, and the resultant interference effect. Other possibilities may be due to the optoelastic effect or the optical uniaxial anisotropy of the LiNbO$_{3}$ single crystal.

\section{M\lowercase{agneto-optical} K\lowercase{err rotation due to} B\lowercase{arnett field}}\label{sec:discussion_moke}
We shall now examine a possible alternative mechanism that would bring about the polarization rotation due to SAW, that is, the magneto-optical Kerr effect induced by the electron spin polarization. Here, we shall only examine the polarization rotation phenomena for the light impinging on the Al/LiNbO$_{3}$ region, in particular, where the spin polarization is established by the effective magnetic field (the so-called \textit{Barnett field}) associated with the vortex field~\cite{M2013,IMM2014,K2017,K2020} associated with the SAW.

To evaluate the Kerr rotation angle due to the spin polarization, we need to know two values: the strength of the Barnett field, $\vector{B}$, and the Kerr rotation angle per unit magnetic field, $K$, for Al. First, let us evaluate the Barnett field $\vector{B}$. The Barnett field $\vector{B}$ for electron spins is given by $\vector{B}=\boldsymbol{\Omega}/2\gamma$, where $\boldsymbol{\Omega}$ is the vortex field accompanying with the SAW and $\gamma=2.8\times10^{10}\,\rm{Hz/T}$ is the gyromagnetic ratio of the electron spin~\cite{M2013,IMM2014,K2017,K2020}. Within the plane-wave approximation, the displacement field $\vector{u}$ has only the $x$ and $z$ components. Thus, the vortex field $\boldsymbol{\Omega}=\vector{\nabla} \times \dot{\vector{u}}$ has the $y$ component only. With the frequency of the SAW, $\omega_{\mathrm{SAW}}/2\pi \sim 86$~MHz, the amplitude of $\vector{u}$ that is determined by the power of the SAW ($-13$~dBm)~\cite{SK2015}, and the elastic parameters of the $\linb$ substrate~\cite{AM2009} listed in Table \ref{table:LN}, the strength of the vortex field $\Omega_y$ in the Al film on the LiNbO$_{3}$ substrate is estimated as $\Omega_{y}/2 \pi \sim 310$~Hz. Note that we treat $\linb$ as an isotropic material for simplicity. We then obtain the $y$ component of the Barnett field $B_y=\Omega_y/2\gamma\sim 5.4\times10^{-9}\,\rm{T}$. The field $B_y$ would polarize the electron spins in the Al film along the $y$-axis.

\begin{table}[h]
  \begin{center}
    \caption{Elastic parameters of $\linb$ \cite{AM2009}}
    \begin{tabular}{|l|c|} \hline
      Parameter & Value  \\ \hline \hline
      Mass density & $4.63\times10^3\,\rm{kg/m^3}$  \\ \hline 
      Bulk modulus & $103\,\rm{GPa}$  \\ \hline
      Shear modulus & $72\,\rm{GPa}$   \\ \hline
    \end{tabular}
  \label{table:LN}
  \end{center}
\end{table}

As for the Kerr angle $K$, we shall resort to the experimentally observed value. The polar Kerr effect was observed for Al with the value of $8.7\times10^{-6}\,\rm{rad/T}$ as the rotation angle per unit magnetic field~\cite{SM1964}. Since the polarization rotation angles of the longitudinal and polar Kerr effect for same material are the same order \cite{YS1996}, we can use this value for our estimate of $K$. The expected angle of polarization rotation due to the magneto-optical Kerr effect is then given by $\theta = B_{y}K$, which amounts to $\theta\sim 10^{-13}\,\rm{rad}$. 

Since the value is many orders of magnitude smaller than the measured value of the order of $10^{-6}\,\rm{rad}$, the magneto-optical Kerr effect should not be the main cause for the observed polarization rotation.

\end{widetext}

\end{document}